\documentclass[a4paper,11pt]{article}
\usepackage{caption}

\usepackage[utf8]{inputenc}
\usepackage{a4wide}
\usepackage{tabularx} % extra features for tabular environment
\usepackage{authblk}
\usepackage{color,xcolor}
\usepackage{newunicodechar}
\newunicodechar{，}{,}
\usepackage{amsmath}  % improve math presentation
\usepackage{amssymb}
\usepackage{graphicx} % takes care of graphic including machinery
\usepackage{subcaption} % 插入子图并排显示
\usepackage{cite} % takes care of citations
\usepackage[final]{hyperref} % adds hyper links inside the generated pdf file
\usepackage{amsmath} % 确保加载 amsmath 宏包
\usepackage{physics}
\usepackage{booktabs}
\def\linkcolor{cyan!70!black}
\hypersetup{
	colorlinks=true,       % false: boxed links; true: colored links
	linkcolor=\linkcolor,        % color of internal links
	citecolor=\linkcolor,        % color of links to bibliography
	filecolor=\linkcolor,     % color of file links
	urlcolor=\linkcolor         
}
\usepackage{blindtext}
\usepackage{cleveref}
\usepackage{appendix}
\definecolor{orange}{rgb}{1,0.5,0}

\newcommand{\JY}[1]{{\color{blue}[JY: #1]}}

\newcommand{\LTW}[1]{{\color{brown}[LTW: #1]}}

\begin{document}

\title{
Primordial Black Hole and Gravitational Wave by \\
Peaked Cosmic Perturbations from Axion Curvaton 
%PBH Formation and Gravitational Wave from Modified Type I Axion Curvaton Model
}
\author[1,2]{Keisuke Harigaya\thanks{kharigaya@uchicago.edu}}
\author[1]{Lian-Tao Wang\thanks{liantaow@uchicago.edu}}
\author[1,3,4]{Jiangyi Yi\thanks{jiangyiy@mit.edu}}

\affil[1]{\emph{\large Enrico Fermi Institute, Kavli Institute for Cosmological Physics, Leinweber Institute for Theoretical Physics, Department of Physics, The University of Chicago, Chicago, IL 60637, USA}}
\affil[2]{\emph{\large Kavli Institute for the Physics and Mathematics of the Universe (WPI),
The University of Tokyo Institutes for Advanced Study,
The University of Tokyo, Kashiwa, Chiba 277-8583, Japan}}
\affil[3]{\emph{\large School of Physics, Nankai University, Tianjin 300071, China}}
\affil[4]{\emph{\large Center for Theoretical Physics- a Leinweber Institute,
Massachusetts Institute of Technology, Cambridge, MA 02139 USA}}
\date{\today}

\maketitle

\begin{abstract}
We demonstrate that an axion field can develop perturbations peaked at an intermediate wavenumber through its inflationary dynamics. We compute the resultant spectrum of gravitational waves (GWs) and primordial black holes (PBHs) generated through the curvaton mechanism. The GW spectrum exhibits a triangular shape, with UV and IR slopes determined by the axion's inflationary evolution and treatable as free parameters. In computing the PBH abundance, we properly account for the nonlinear relation between curvature and density perturbations and the primordial non-Gaussianity generated by the curvaton. The PBHs can account for all of dark matter, accompanied by blue-tilted isocurvature perturbations at small scales. The IR slope of the GW spectrum is anti-correlated with the magnitude of the dark matter isocurvature perturbations, offering a multi-messenger test of the model.
\end{abstract}

\newpage
\tableofcontents

\section{Introduction}

Cosmic perturbations are precisely measured on large scales, $10^{-4}~\text{Mpc}^{-1} \lesssim k \lesssim 1~\text{Mpc}^{-1}$, where they are nearly scale-invariant, Gaussian, and adiabatic~\cite{Planck:2018vyg}. Smaller scales, however, remain largely unexplored.

A variety of particle-physics models can enhance small-scale perturbations while leaving large-scale perturbations intact. For example, scalar fields with masses comparable to the Hubble scale during inflation can produce blue-tilted curvature perturbations~\cite{Kawasaki:2013xsa,Ebadi:2023xhq} via the curvaton mechanism~\cite{Mollerach:1989hu, Linde:1996gt, Enqvist:2001zp, Lyth:2001nq, Moroi:2001ct}. Axion fields can similarly generate blue-tilted perturbations~\cite{Kasuya:2009up}. In such models, large curvature perturbations at small scales can produce observable gravitational waves (GWs)~\cite{Ananda:2006af,Baumann:2007zm} and primordial black holes (PBHs)~\cite{Carr:1974nx}.

In this paper, we show that an axion field in a supersymmetric model can develop field fluctuations with a spectrum peaked at a characteristic wavenumber. The infrared (IR) and ultraviolet (UV) slopes of the peaked spectrum are controlled by the Hubble-induced masses of the symmetry-breaking fields and may be treated as free parameters. A rotating axion field can also be the QCD axion that solves the strong CP problem. 

These field perturbations source energy-density fluctuations in the oscillating or rotating axion field, which is subsequently converted into radiation, generating adiabatic perturbations.
Provided the mode at the spectral peak reenters the horizon after this conversion, the peak in the field fluctuations maps directly onto a peak in the adiabatic perturbation spectrum. We compute the resulting GW and PBH abundances from these peaked adiabatic perturbations, accounting for the effects of non-Gaussianity.

The GW spectrum inherits a peak from the adiabatic perturbations, and depending on the UV and IR slopes, the signal may be detectable across multiple GW observatories. In particular, our setup can accommodate the GWs observed by NANOGrav~\cite{NANOGrav:2023gor,NANOGrav:2023hde,NANOGrav:2023hvm}.

PBHs with masses up to the horizon mass at the spectral peak are also produced. Due to the exponential sensitivity of the PBH abundance to the amplitude of the perturbations, the PBH mass function is sharply suppressed away from the peak. Due to the peaked spectrum, PBHs can constitute all of dark matter without violating evaporation~\cite{Clark:2016nst,Mittal:2021egv,Laha:2020ivk,Berteaud:2022tws,Boudaud:2018hqb,DeRocco:2019fjq,Carr:2009jm} and microlensing constraints~\cite{Niikura:2017zjd,EROS-2:2006ryy,Niikura:2019kqi,Oguri:2017ock}, which place stringent upper bounds on the abundance of light PBHs.

The PBH dark matter carries isocurvature perturbations on large scales, inherited from the axion field fluctuations. The amplitude of these isocurvature perturbations is anti-correlated with the IR slope of the GW spectrum. This  relationship establishes a falsifiable multi-messenger test: a precise measurement of the GW IR slope directly predicts the isocurvature amplitude. Consequently, this provides a definitive and testable target for future small-scale matter power spectrum observations, such as 21-cm cosmology~\cite{deKruijf:2024voc}, linking GW searches with the large-scale structure of the universe.

This paper is organized as follows. In Sec.~\ref{sec:model}, we present the axion curvaton model and derive the peaked field fluctuation spectrum. In Secs.~\ref{sec:PBH} and~\ref{sec:GW}, we describe the computation of the PBH and GW abundances. Results are presented in Sec.~\ref{sec:results}, and we summarize in Sec.~\ref{sec:summary}.

\section{Axion Curvaton Model}
\label{sec:model}

We begin by outlining the cosmological scenario of our model. During the inflation, complex scalar fields charged under a $U(1)$ symmetry move on their moduli space driven by their Hubble induced masses. The dynamics leads to  fluctuations of an axion field that are peaked at a certain wavenumber.
The axion field later nearly dominates the universe and decays into radiation. Because the axion energy density depends on its initial field value, curvature perturbations with a peaked spectrum are created. The IR and UV slopes are determined by the magnitudes of the Hubble-induced masses of the $U(1)$ charged scalars. This peaked spectrum subsequently leads to the production of PBHs and scalar-induced GWs, which are the main subjects of this work.

In the following, we present the explicit supersymmetric model and analyze its dynamics. The Hubble parameter $H_{\mathrm{inf}}$ during inflation is assumed to be a constant number with unspecified inflation model.

\subsection{Potential and Dynamics}
\label{sec:potential_dynamics}
The superpotential of the axion curvaton model is given by~\cite{Kasuya:1996ns}
\begin{equation}
W = hS(\Phi \bar{\Phi} - f^2),
\end{equation}
where $\Phi$, $\bar{\Phi}$, and $S$ are chiral superfields, $f$ is a characteristic energy scale, and $h$ is a dimensionless coupling constant. The corresponding supersymmetric scalar potential reads
\begin{equation}
V = h^2|\Phi \bar{\Phi} - f^2|^2 + h^2|S|^2 \left(|\Phi|^2 + |\bar{\Phi}|^2\right).
\end{equation}
The potential exhibits a flat direction determined by the $F$-term conditions:
\begin{equation}
\Phi \bar{\Phi} = f^2,\quad S = 0. \label{eq:flat_condition}
\end{equation}
%We assume that this condition is satisfied throughout the evolution of the fields. 

Including supergravity effects, Hubble-induced mass terms can arise from the K\"ahler potential~\cite{Dine:1995kz}. They are of the form 
\begin{equation}
V_H = c_1 H^2 |\Phi|^2 + c_2 H^2 |\bar{\Phi}|^2 + c_S H^2 |S|^2,
\end{equation}
where $c_1$, $c_2$, and $c_S$ are dimensionless constants of order unity, which may take either sign. In the present analysis, we consider the case $c_1 > 0$ and $c_2 < 0$. We further assume that the initial field configuration satisfies $|\Phi| \gg |\bar{\Phi}|$.

\if0
\LTW{For the constraint in equation 3 to be valid, the mode transverse to the trajectory should be heavier than the masses in equation 4. Is the always satisfied in our case (with $H=3f$ in figure 1)?}
\JY{The mode transverse to the flat direction has mass

$$
m_\perp^2
=
h^2\left(|\Phi|^2+|\bar\Phi|^2\right)
=
\frac{h^2}{2}\varphi^2.
$$

Since \(\varphi\geq 2f\) along \(\Phi\bar\Phi=f^2\), its minimum mass is

$$
m_{\perp,\min}^2=2h^2f^2.
$$

Thus, imposing Eq.~(3) requires

$$
2h^2f^2
\gtrsim
H^2\max\!\left(c_1,|c_2|\right).
$$

For \(H=3f\),

$$
\frac{m_{\perp,\min}^2}{H^2}
=
\frac{2h^2}{9}.
$$

For the parameter sets used in the tables, \(c_1\lesssim1.74\) and \(|c_2|\lesssim1\), so \(h\gtrsim2.8\) makes the transverse mode heavier than the Hubble-induced modes. We may therefore take

$$
h\simeq3\text{--}5.
$$

This remains perturbative since

$$
\frac{h^2}{16\pi^2}\simeq0.06\text{--}0.16<1.
$$

For the illustrative \(c_2=-4\) curve in Fig.~1, the corresponding condition is \(h\gtrsim4.24\).
}
\fi

It is convenient to decompose the fields into radial and angular components:
\begin{equation}
\Phi = \frac{1}{\sqrt{2}} \varphi  e^{i\theta_{+}},\quad
\bar{\Phi} = \frac{1}{\sqrt{2}} \bar{\varphi} e^{i\theta_{-}}. \label{eq:decomposition}
\end{equation}
The axion curvaton is identified with a combination of these degrees of freedom. In particular, the canonical axion field is defined as
\begin{equation}
\tilde{\sigma} \equiv \sqrt{\varphi^2 + \bar{\varphi}^2} \, \theta \equiv \phi \, \theta,
\end{equation}
where $\phi$ denotes the total radial field and $\theta=\theta_+=-\theta_-$.

Imposing the flatness condition Eq.~(\ref{eq:flat_condition}) together with the decomposition Eq.~(\ref{eq:decomposition}), one obtains the effective Lagrangian for the radial field $\varphi$:
\begin{equation}
\mathcal{L}_{\mathrm{eff}}(\varphi)
= \frac{1}{2} \left( 1 + \frac{4f^4}{\varphi^4} \right)(\partial_\mu \varphi)^2
- V(\varphi),
\end{equation}
with the potential
\begin{equation}
V(\varphi)
= \frac{1}{2} c_1 H^2 \varphi^2
+ \frac{2 c_2 H^2 f^4}{\varphi^2}. \label{eq:V_eff}
\end{equation}
Analogously, the effective Lagrangian for the angular field $\theta$ takes the form
\begin{equation}
\mathcal{L}_{\mathrm{eff}}(\theta)
= \frac{1}{2} \frac{\varphi^4 + 4f^4}{\varphi^2} (\partial_\mu \theta)^2.
\end{equation}

\begin{figure}[t!]
    \centering
    \begin{subfigure}[b]{0.48\textwidth}
        \centering
        \includegraphics[width=\textwidth]{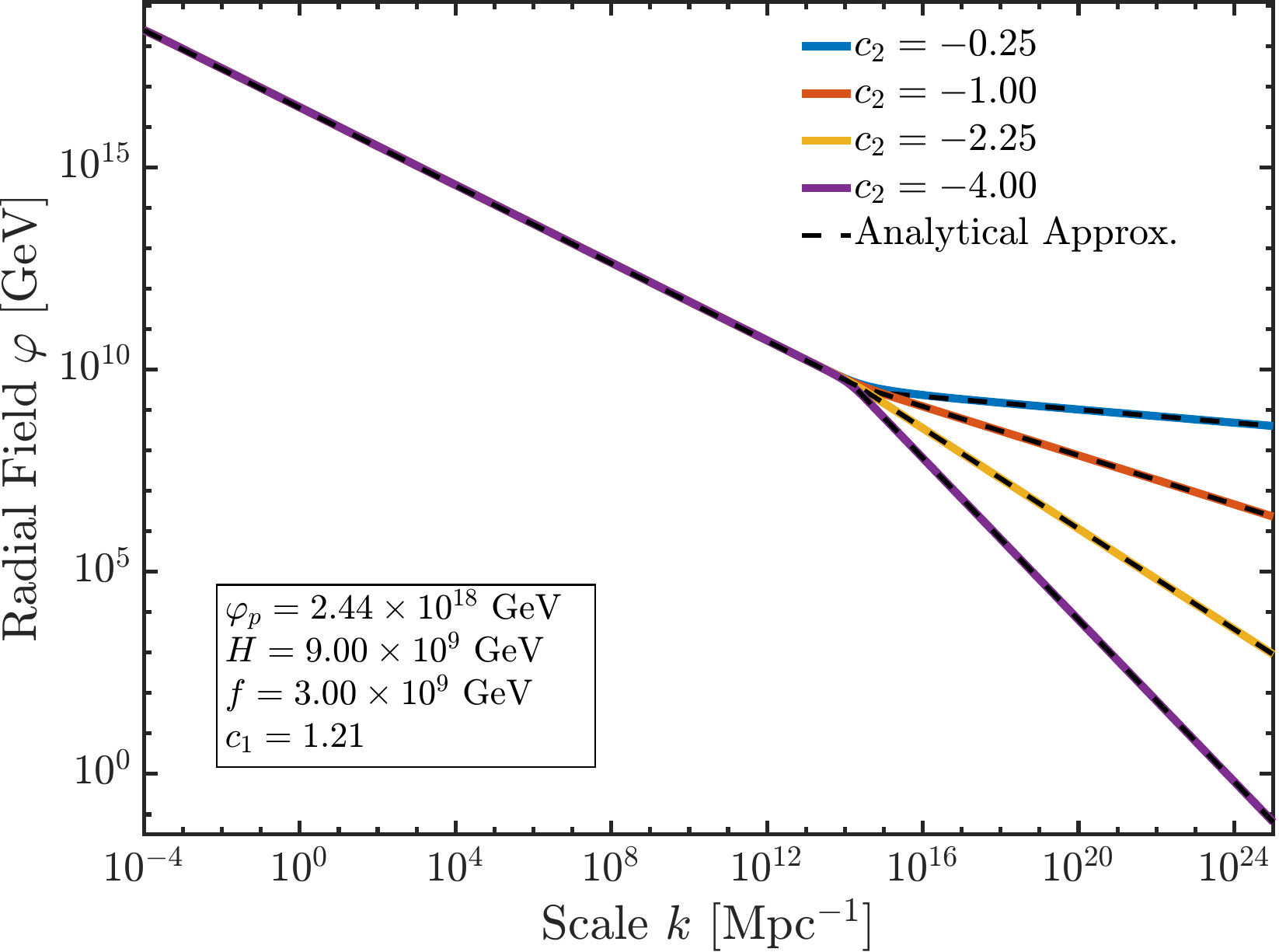}  
        \caption{Radial field $\varphi$.}
    \end{subfigure}
    \hfill
    \begin{subfigure}[b]{0.48\textwidth}
        \centering
        \includegraphics[width=\textwidth]{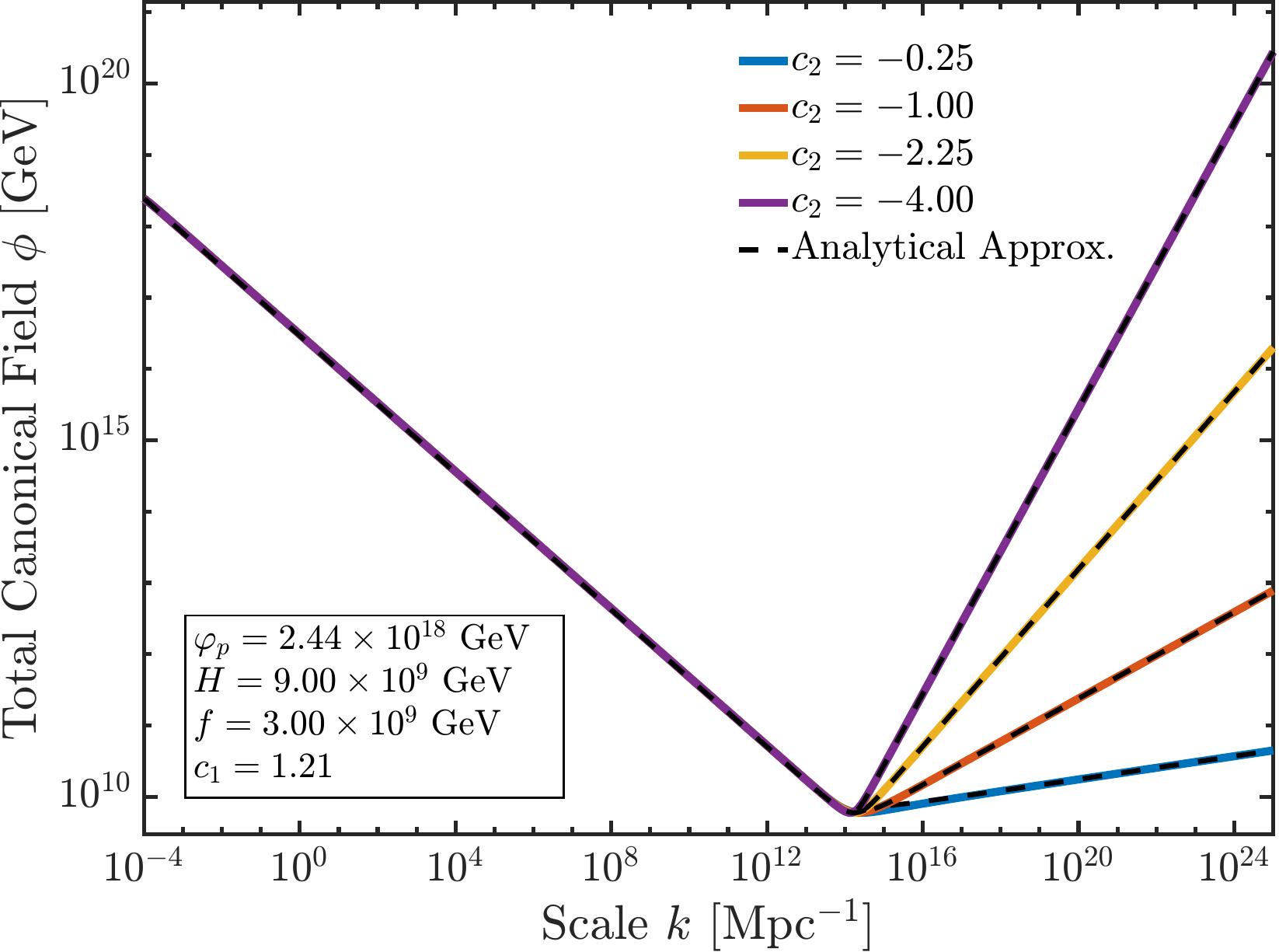} 
        \caption{Total canonical field $\phi = \sqrt{\varphi^2 + 4f^4/\varphi^2}$.}
    \end{subfigure}   
    \caption{Evolution of the fields at horizon exit as a function of the comoving wavenumber $k$.  
    \textbf{Left panel:} Radial field $\varphi$. Colored solid curves are full numerical solutions for different $c_2$; thin black dashed lines show piecewise analytical approximations. Before reaching $f$, $\varphi \propto k^{-\lambda_1}$ (positive Hubble mass dominates); after $\varphi \lesssim f$, the negative mass term drives $\varphi \propto k^{\lambda_2}$, with $\lambda_2$ depending on $c_2$.  
    \textbf{Right panel:} Total canonical field $\phi$, which directly controls the amplitude of the curvaton fluctuation via $\mathcal{P}_{\delta\theta}^{1/2} = H_{\mathrm{inf}}/(2\pi\phi)$ (see Eq.~\eqref{eq:P_delta_theta}). Its V-shaped curve features a minimum at $k = k_{\mathrm{peak}}$, corresponding to the peak of the curvature power spectrum. The same analytical approximations (black dashes) are shown for $\phi$ as well, which correctly capture the full scale-dependent evolution.} 
    \label{fig:phi_evolution}
\end{figure}

The equation of motion for $\varphi$ in a Friedmann–Robertson–Walker (FRW) background is
\begin{equation}
\left( 1 + \frac{4f^4}{\varphi^4} \right)\ddot{\varphi}
- 8\frac{f^4}{\varphi^5} \dot{\varphi}^2
+ 3H \left( 1 + \frac{4f^4}{\varphi^4} \right)\dot{\varphi}
= -c_1 H^2 \varphi + 4c_2 H^2 f^4 \varphi^{-3},
\end{equation}
where the back reaction from $\theta$ has been neglected, which is a justified approximation in the regime of interest. Owing to the negative mass contribution from $\bar{\Phi}$ (i.e., $c_2<0$), the field $\varphi$ rolls monotonically towards smaller values. The variation of $\varphi$ at the time of horizon exit for different comoving scales $k$ is illustrated in Fig.~\ref{fig:phi_evolution}. In this figure, we fix parameter $c_1$ and vary $c_2$ to demonstrate the field's scale-dependent dynamics. The black dashed line represents an analytical approximation. In the early stage ($\varphi \gg \sqrt{2}f$) the $c_2$ term is negligible. The equation of motion simplifies to $\ddot{\varphi} + 3H\dot{\varphi} + c_1 H^2 \varphi \simeq 0$, yielding a power-law evolution $\varphi \propto k^{-\lambda_1}$ with the index $\lambda_1 = \frac{3}{2} - \frac{3}{2}\sqrt{1 - \frac{4}{9}c_1}$. As shown, the numerical solutions perfectly match this approximation at large scales (small $k$). 

As $\varphi$ rolls down and falls below the symmetry-breaking scale $f$, the negative mass-squared term proportional to $c_2$ becomes dominant. In this regime, we can use the approximation where $\bar{\Phi}$ is the canonical field. Specifically, the canonical radial field becomes $\bar{\varphi} \approx 2f^2/\varphi$, and its equation of motion simplifies to $\ddot{\bar{\varphi}} + 3H\dot{\bar{\varphi}} - |c_2|H^2\bar{\varphi} \simeq 0$. This yields a growing mode for $\bar{\varphi}$, meaning that $\varphi$ transitions to a new power-law decay $\varphi \propto k^{\lambda_2}$ with the index $\lambda_2 = \frac{3}{2} - \frac{3}{2}\sqrt{1 + \frac{4}{9}|c_2|}$. As illustrated in Fig.~\ref{fig:phi_evolution}, this late-stage analytical approximation captures the slopes for different $c_2$ values after the turning point. The evolution of $\varphi$ is distinctive from that for $c_2>0$, where $\varphi$ nearly stops evolving at $\varphi \sim f$.

To summarize, the dynamics of $\varphi$ is governed by the parameters $c_1$ and $c_2$ in distinct regimes: when $\varphi \gtrsim \sqrt{2}f$, the term proportional to $c_1$ dominates, whereas for $\varphi \lesssim \sqrt{2}f$, the term proportional to $c_2$ becomes important. This transition leads to a scale-dependent evolution of the field, which ultimately generates a peaked spectrum of curvature perturbations, enabling the production of PBHs as the entirety of dark matter.

In our analysis, we integrate out the direction transverse to the flat direction $\Phi \bar{\Phi}=f^2$, which requires that the transverse direction be heavier than the flat direction.
The mode transverse to the flat direction has a mass
\begin{equation}
    m_\perp^2
=
h^2\left(|\Phi|^2+|\bar\Phi|^2\right)
=
\frac{h^2}{2}\varphi^2.
\end{equation}
Since \(\varphi\geq 2f\) along \(\Phi\bar\Phi=f^2\), its minimum  is
\begin{equation}
    m_{\perp,\min}^2=2h^2f^2.
\end{equation}
Thus, integrating out the transverse direction requires
\begin{equation}
    2h^2f^2
\gtrsim
H^2\max\!\left(c_1,|c_2|\right).
\end{equation}
The maximal $H$ we will consider is $3f$, for which
\begin{equation}
    h > 2\times \sqrt{\max\!\left(c_1,|c_2|\right)}
\end{equation}
is required.
For $c_1,|c_2| = O(1)$, the required $h$ remains perturbative around the scale $f$. $h$ may become non-perturbative at higher scales, for which case some UV completion is required, which can be achieved by a theory with deformed moduli constraints~\cite{Seiberg:1994bz}. For smaller $H/f$, $h$ may remain perturbative up to the Planck scale.

\subsection{Curvature perturbation}
\label{sec:curvper}
We now consider the perturbations associated with the phase direction of the axion curvaton. During inflation, quantum fluctuations generate perturbations in the curvaton field. At horizon exit of a comoving scale $k$, the fluctuation in the angular field $\theta$ has an amplitude of order $H_{\mathrm{inf}}/(2\pi\phi(k))$, where $\phi(k)$ denotes the value of the radial field at that moment. The corresponding power spectrum is
\begin{equation}
\mathcal{P}_{\delta\theta}^{1/2}(k) = \frac{H_{\mathrm{inf}}}{2\pi\phi(k)}. \label{eq:P_delta_theta}
\end{equation}
These fluctuations remain conserved on super-horizon scales as long as the mass of $\theta$ is smaller than the Hubble scale.

The $U(1)$ symmetry is assumed to be explicitly broken, generating a potential for the curvaton. Note that the decay constant governing this potential is determined by the vacuum expectation value (VEV) of the radial field after inflation, denoted as $F$, which may be equal to or significantly larger than $f$. The potential takes the form
\begin{equation}
V_{\sigma} = \Lambda^{4}\left[1 - \cos\left(\frac{\sigma}{F}\right)\right] 
\simeq \frac{1}{2}m_{\sigma}^{2}\sigma^{2}, \label{eq:curvaton_potential}
\end{equation}
where the curvaton mass is $m_{\sigma} = \Lambda^{2}/F$. The quadratic approximation in Eq.~(\ref{eq:curvaton_potential}) is valid when $\sigma/F \ll 1$, which we assume throughout for simplicity. When the Hubble parameter drops below $m_{\sigma}$, the curvaton begins to oscillate with an initial amplitude $\sigma_i = F\theta$, where $\theta$ is the misalignment angle. As we will discuss in Section \ref{subsec:PBH_isocurvature}, a successful PBH dark matter scenario often requires the vacuum decay constant to be dynamically enhanced ($F \gg f$) to ensure curvaton energy domination while satisfying isocurvature constraints.

The energy density of the oscillating curvaton scales as $\rho_{\sigma} \propto a^{-3}$, while the radiation background scales as $\rho_r \propto a^{-4}$. This leads to a growing ratio $r \equiv \rho_{\sigma}/\rho_r \propto a$ until the curvaton decays. In this work, we focus on scenarios where the curvaton (nearly) dominates the energy density at decay, i.e., $r_{D} \equiv r(\eta_{\mathrm{dec}}) \gtrsim 1$, where $\eta_{\mathrm{dec}}$ is the conformal time at curvaton decay. As discussed in Ref.~\cite{Inomata:2023drn}, this regime leads to a negative non-Gaussianity parameter $f_{\mathrm{NL}}$, which suppresses primordial black hole formation.

The fraction $r_D$ is given by
\begin{equation}
r_D =
\begin{cases}
\frac{1}{6}\left(\frac{F\theta}{M_P}\right)^2\frac{T_R}{T_{\mathrm{dec}}}\quad\mathrm{for}\quad m_\sigma\geq\Gamma_I, \\
\frac{1}{6}\left(\frac{F\theta}{M_P}\right)^2\frac{T_{\mathrm{osc}}}{T_{\mathrm{dec}}}\quad\mathrm{for}\quad m_\sigma\leq\Gamma_I,
\end{cases}
\label{eq:rD}
\end{equation}
where $T_R$, $T_{\mathrm{dec}}$, and $T_{\mathrm{osc}}$ denote the temperatures of the thermal bath at the completion of reheating, the curvaton decay, and the onset of curvaton oscillation, respectively, and $\Gamma_I$ is the decay rate of the inflaton. The condition $m_\sigma \geq \Gamma_I$ ($m_\sigma \leq \Gamma_I$) corresponds to the scenario where the curvaton begins to oscillate before (after) the completion of reheating.

We assume that the modes relevant to PBH formation re-enter the horizon after the curvaton decays. Then the separate universe approximation is valid and we can treat $r$ as a constant when evaluating the curvature perturbations for these scales. This simplifies the analysis considerably.

At leading order, the perturbation in the curvaton energy density is related to the field fluctuation by
\begin{equation}
\frac{\delta\rho_{\sigma}}{\rho_{\sigma}} = \frac{2\delta\sigma}{\sigma} = \frac{2\delta\theta}{\theta}. \label{eq:deltarho_curv}
\end{equation}
Evaluated on the spatially flat slicing, where the primordial radiation perturbations are assumed to be subdominant, the gauge-invariant isocurvature perturbation induced by the curvaton is approximately
\begin{equation}
\mathcal{S} \equiv \frac{\delta\rho_{\sigma}}{\rho_{\sigma}} - \frac{3}{4}\frac{\delta\rho_r}{\rho_r} \simeq \frac{\delta\rho_{\sigma}}{\rho_{\sigma}}.
\end{equation}
As we will see in Section \ref{subsec:PBH_isocurvature}, this primordial isocurvature perturbation $\mathcal{S}$ seeds the PBH isocurvature perturbations ($\mathcal{S}_{\mathrm{PBH}}$), which subjects the model to stringent observational constraints from the cosmic microwave background (CMB).
The corresponding power spectrum follows from Eq.~(\ref{eq:P_delta_theta}),
\begin{equation}
\mathcal{P}_{\mathcal{S}}^{1/2}(k) = \frac{H_{\mathrm{inf}}}{\pi\phi(k)\theta}. \label{eq:Ps}
\end{equation}

The curvature perturbation $\zeta$ at linear order (the Gaussian part) is given by 
\begin{equation}
\zeta_{\rm G} = -H\frac{\delta\rho}{\dot{\rho}},
\end{equation}
where the right-hand side is evaluated on the flat slicing of the spacetime. Using $\rho = \rho_r + \rho_{\sigma}$ and the Friedmann equations, this becomes
\begin{equation}
\zeta_{\rm G} = \frac{\delta\rho_r + \delta\rho_{\sigma}}{4\rho_r + 3\rho_\sigma} 
\simeq \frac{\delta\rho_\sigma}{4\rho_r + 3\rho_\sigma},
\end{equation}
where we have neglected $\delta\rho_r$ relative to $\delta\rho_\sigma$. Substituting Eq.~(\ref{eq:deltarho_curv}), we obtain the Gaussian curvature perturbation power spectrum sourced by the curvaton,
\begin{equation}
\mathcal{P}_{\zeta}(k,\eta) 
= \left( \frac{2}{3}r_{dec}(\eta) \right)^2 
\mathcal{P}_{\theta}(k) 
= \left( \frac{2}{3}r_{dec}(\eta) \right)^2 
\left( \frac{H_{\mathrm{inf}}}{2\pi\phi(k)\theta} \right)^2, \label{eq:Pzeta_curv_eta}
\end{equation}
where we have defined
\begin{equation}
r_{\mathrm{dec}}(\eta) \equiv \frac{3\rho_\sigma(\eta)}{4\rho_r(\eta) + 3\rho_\sigma(\eta)}=\frac{3r(\eta)}{4 + 3r(\eta)} \label{eq:r_definition}
\end{equation}
for later convenience.
As noted above, for modes that re-enter after the curvaton decays, we can evaluate the curvature perturbation at $\eta > \eta_{\mathrm{dec}}$ with constant $r = r_{\mathrm{dec}}$. Thus,
\begin{equation}
\mathcal{P}_{\zeta}(k) \equiv \mathcal{P}_{\zeta}(k,\eta > \eta_{\mathrm{dec}}) 
= \left( \frac{2}{3}r_{\mathrm{dec}} \right)^2 
\left( \frac{H_{\mathrm{inf}}}{2\pi\phi(k)\theta} \right)^2. \label{eq:Pzeta_curv_final}
\end{equation}

The scale dependence of $\mathcal{P}_{\zeta}(k)$ is entirely determined by $\phi(k)$, the radial field value at horizon exit. As discussed in Sec.~\ref{sec:potential_dynamics}, $\phi(k)$ evolves during inflation in a non-monotonic manner: initially decreasing with $k$, reaching a minimum, and then increasing again for larger $k$. This behavior translates into a corresponding peak in the curvature power spectrum, as opposed to the plateau-like spectrum in the standard axion-like curvaton model. Specifically, for scales that exit the horizon when $\phi(k)$ is decreasing, $\mathcal{P}_{\zeta}(k)$ increases with $k$ (blue-tilted). When $\phi(k)$ reaches its minimum, the power spectrum reaches a maximum. For even smaller scales that exit when $\phi(k)$ is increasing, $\mathcal{P}_{\zeta}(k)$ decreases with $k$, producing a red-tilted segment. Consequently, the overall power spectrum exhibits a peak at a characteristic scale $k_{\mathrm{peak}}$, which corresponds to the scale that  exits the horizon during inflation when $\phi$ is at its minimum value.

This peaked spectrum is crucial for generating a narrow mass distribution of PBHs, which has the possibility to match the LIGO or LISA observations~\cite{Kawasaki:2012wr,Ando:2017veq,Inomata:2020xad,Ferrante:2023bgz}.

We have discussed the case where the axion oscillates around the minimum of the potential. The axion field may instead rotate in the field space~\cite{Co:2019wyp,Co:2019jts}, initiated by the Affleck-Dine mechanism~\cite{Affleck:1984fy}.
After inflation, $\bar{\Phi}$ is assumed to have a negative Hubble induced mass term to take on a large field value. When the vacuum mass of $\bar{\Phi}$, which is assumed to be positive, exceeds the Hubble scale, $\bar{\Phi}$ begins to oscillate. The $U(1)$ symmetry is explicitly broken by a higher-dimensional potential term, which kicks the axion field to the angular direction to initiate rotations. 
Because of the approximate $U(1)$ charge conservation, the rotation is stable against dissipation even if it couples to the thermal bath.
The radial value of $\bar{\Phi}$ decreases by the Hubble expansion.
Since the radial direction has a quadratic potential, the energy density of the rotation scales as matter and eventually dominates the universe. If $U(1)$ symmetry is explicitly broken by some interaction (e.g., sphaleron processes), the $U(1)$ charge can be washed out and the universe can be reheated by the axion rotation.
This curvaton scenario has similar phenomenology as the oscillating axion; see~\cite{Co:2022qpr} for details.

Rotating axion may generate the baryon asymmetry of the universe via axiogenesis~\cite{Co:2019wyp,Co:2020jtv,Harigaya:2021txz,Chakraborty:2021fkp,Kawamura:2021xpu,Co:2021qgl,Barnes:2022ren,Barnes:2024jap}. Also, since the $U(1)$-violating potential is effective only at large field values of $\bar{\Phi}$ and may be negligible at the vacuum, the axion field here may be the QCD axion~\cite{Peccei:1977hh,Peccei:1977ur,Weinberg:1977ma, Wilczek:1977pj} that solves the strong CP problem.

\subsection{Non-Gaussianity: Full Order}

In this section, we employ the $\delta N$ formalism~\cite{Sasaki:1995aw,Wands:2000dp,Lyth:2004gb} to derive the complete nonlinear relation between the full curvature perturbation $\zeta$ and its Gaussian component $\zeta_{\rm G}$. This approach extends beyond the second-order calculation typically presented in the literature, allowing us to capture all higher-order non-Gaussian (NG) contributions that are crucial for accurate evaluation of the PBH abundance, especially in the regime where the curvaton dominates at decay ($r_{D}>1$). Here and for the rest of this paper, we compute the PBH abundance for the oscillating curvaton scenario, but qualitative features are expected to be the same for the rotating curvaton scenario.

Let $t_{\mathrm{osc}}$ and $t_{\mathrm{dec}}$ denote the times when the curvaton begins oscillation and when it decays, respectively. The Friedmann equations at these moments read
\begin{equation}\begin{aligned}
 & 3m_{\sigma}^{2}M_{P}^{2}=\rho_{r,\mathrm{osc}}+\rho_{\sigma,\mathrm{osc}}\simeq\rho_{r,\mathrm{osc}}, \\
 & 3\Gamma_{\sigma}^{2}M_{P}^{2}=\rho_{r,\mathrm{dec}}+\rho_{\sigma,\mathrm{dec}}=\left(\frac{a_{\mathrm{osc}}}{a_{\mathrm{dec}}}\right)^{4}\rho_{r,\mathrm{osc}}+\left(\frac{a_{\mathrm{osc}}}{a_{\mathrm{dec}}}\right)^{3}\rho_{\sigma,\mathrm{osc}},
\end{aligned}\label{eq:friedmann_osc_dec}\end{equation}
where the subscripts “osc” and “dec” indicate evaluation at $t=t_{\mathrm{osc}}$ and $t=t_{\mathrm{dec}}$, respectively. The energy density of the oscillating curvaton scales as matter, $\rho_\sigma \propto a^{-3}$, while the radiation density scales as $\rho_r \propto a^{-4}$.\footnote{If $m_\sigma>\Gamma_I$, the curvaton
starts oscillating before inflaton reheating. Before reheating, both
the curvaton and the inflaton condensate redshift as $a^{-3}$, giving
$\delta N_{\rm osc\to R}=0$ between uniform-total-density hypersurfaces
in the sudden-reheating approximation. The radiation-plus-curvaton
evolution used below then begins at reheating. We neglect corrections
to the curvaton energy fraction at that time.}   %\LTW{Does this discussion cover the case $m_\sigma > \Gamma_I$, where the inflaton decay will have some entropy injection?} \KH{Apparently not, but the result should remain the same since $\delta N$ from the beginning of the oscillation to the reheating by the inflation is zero. We can say that the result remains the same for that case.}\JY{I tried to add it to footnote}

We define the perturbed and unperturbed number of $e$-folds between $t_{\mathrm{osc}}$ and $t_{\mathrm{dec}}$ as
\begin{equation}
e^{N}\equiv\frac{a_{\mathrm{dec}}}{a_{\mathrm{osc}}},\qquad e^{N_0}\equiv\frac{\bar{a}_{\mathrm{dec}}}{\bar{a}_{\mathrm{osc}}},
\end{equation}
where an over-bar denotes the homogeneous, unperturbed quantity. The curvature perturbation on uniform-density slices is then given by $\zeta = N - N_0$. From Eq.~\eqref{eq:friedmann_osc_dec} we obtain the relation
\begin{equation}
e^{-4N}+e^{-3N}\,\frac{\rho_{\sigma,\mathrm{osc}}}{\rho_{r,\mathrm{osc}}}= \frac{\rho_{t,\mathrm{dec}}}{\rho_{r,\mathrm{osc}}},
\end{equation}
where $\rho_{t,\mathrm{dec}} \equiv \rho_{r,\mathrm{dec}} + \rho_{\sigma,\mathrm{dec}}$. Using the background relations ${\bar \rho_{\sigma,\mathrm{dec}}}={\bar \rho_{\sigma,\mathrm{osc}}}e^{-3N_0}$ and ${\bar \rho_{r,\mathrm{dec}}}={\bar \rho_{r,\mathrm{osc}}}e^{-4N_0}$,  we derive the following algebraic equation for $\zeta$:
\begin{equation}
e^{4\zeta} - \left[ \Omega_{\sigma,\mathrm{dec}}\,\frac{\rho_{\sigma,\mathrm{osc}}}{{\bar \rho_{\sigma,\mathrm{osc}}}} \right] e^{\zeta} + \bigl( \Omega_{\sigma,\mathrm{dec}} - 1 \bigr) = 0,
\end{equation}
where $\Omega_{\sigma,\mathrm{dec}} \equiv \bar{\rho}_{\sigma}/(\bar{\rho}_{r}+\bar{\rho}_{\sigma})$ is the fractional energy density of the curvaton at decay. Within the sudden-decay approximation, the solution of this equation can be expressed in a closed form as \cite{Sasaki:2006kq}
\begin{equation}
\zeta = \ln\!\bigl[ X(r_{\mathrm{dec}},\zeta_{\rm G}) \bigr], \label{eq:zeta_full_nonlinear}
\end{equation}
$X$ is a function of the decay parameter $r_{\mathrm{dec}}=r_{\mathrm{dec}}(\eta_{\mathrm{dec}})$ defined in Eq.~(\ref{eq:r_definition}) and the Gaussian curvature perturbation $\zeta_{\rm G}$.  

For a curvaton with a quadratic potential, $V(\sigma) = \frac12 m_\sigma^2 \sigma^2$, the function $X$ takes the explicit form
\begin{equation}\begin{aligned}
 & X \equiv \frac{\sqrt{K}\Bigl(1+\sqrt{A\,K^{-3/2}-1}\Bigr)}{(3+r_{\mathrm{dec}})^{1/3}}, \\
 & K \equiv \frac12\Bigl( (3+r_{\mathrm{dec}})^{1/3}(r_{\mathrm{dec}}-1)P^{-1/3} + P^{1/3} \Bigr), \\
 & P \equiv A^{2} + \sqrt{A^{4} + (3+r_{\mathrm{dec}})(1-r_{\mathrm{dec}})^{3}}, \\
 & A \equiv \left(1 + \frac{3\zeta_{\rm G}}{2r_{\mathrm{dec}}}\right)^{\!2} r_{\mathrm{dec}}.
 \label{eq:XKPA}
\end{aligned}\end{equation}
If, instead, the curvaton potential is of the cosine type, $V(\sigma) = \Lambda^4[1-\cos(\sigma/F)]$, the expression for $A$ is modified to %\LTW{maybe avoid mention $\cos$ potential since we are using quadratic for the calculation?}\JY{Here I keep this because I tried to emphasize later that current computation on PBH formation is not application on this periodic potential, a real "full" computation is needed}
\begin{equation}
A \equiv \frac{1 - \cos\!\Bigl[ \theta \bigl(1 + \frac{3\zeta_{\rm G}}{2r_{\mathrm{dec}}}\bigr) \Bigr]}{1 - \cos\theta}\; r_{\mathrm{dec}}, \label{eq:A_cosine}
\end{equation}
where $\theta \equiv \sigma_i/F$ is the initial misalignment angle defined in the previous section.

Expanding Equation~\eqref{eq:zeta_full_nonlinear} to second order in $\zeta_{\rm G}$, together with Eqs.~\eqref{eq:XKPA}, we can  extract local non-Gaussianity 
\begin{equation}
\zeta = \zeta_{\rm G} + \frac{3}{5}f_{\mathrm{NL}} \bigl( \zeta_{\rm G}^2 - \langle \zeta_{\rm G}^2 \rangle \bigr) + \mathcal{O}(\zeta_{\rm G}^3), \label{eq:zeta_2nd_order}
\end{equation}
where the nonlinear parameter $f_{\mathrm{NL}}$ is a function of $r_{\mathrm{dec}}$ (and, in the cosine case, of $\theta$). The explicit form of $f_{\mathrm{NL}}$ can be obtained by Taylor expanding Eq.~\eqref{eq:zeta_full_nonlinear}; for the quadratic potential it reduces to the well-known result \cite{Sasaki:2006kq}
\begin{equation}f_{\mathrm{NL}}(r_{\mathrm{dec}})\equiv\frac{5}{3}\left(\frac{3}{4r_{\mathrm{dec}}}-1-\frac{r_{\mathrm{dec}}}{2}\right)\end{equation}
while for the cosine potential $f_{\mathrm{NL}}$ acquires an additional dependence on $\theta$.

The full expression~\eqref{eq:zeta_full_nonlinear} is essential for accurately computing the PBH abundance when non-Gaussianities are large. In particular, the curvature perturbation PDF develops a pronounced non-Gaussian tail, which significantly enhances (or suppresses, depending on the sign of $f_{\mathrm{NL}}$) the probability of forming PBHs relative to the Gaussian estimate. By retaining all orders in $\zeta_{\rm G}$, our formulation enables a precise determination of the PBH mass spectrum. Moreover, the same formalism naturally incorporates the transition between quadratic and cosine potentials, providing a unified framework for studying axion curvaton models beyond the simplest quadratic approximation.

\section{PBH formation}
\label{sec:PBH}

Primordial black holes are formed when the density perturbation, averaged over a Hubble patch, exceeds a critical threshold value $\delta_c$ at the moment the over dense region reenters the Hubble horizon. Here, $\delta_c$ represents the critical density contrast threshold required for an overdense region to overcome internal pressure gradients and successfully undergo gravitational collapse into a black hole. In this work, We only consider the production of PBHs from the modes that enter the horizon after the curvaton decays.

To accurately evaluate the PBH formation and abundance in our model, simple linear estimates are insufficient. Instead, our precise numerical analysis relies on the rigorous compaction function formalism\cite{Ferrante:2022mui}(see also \cite{Gow:2022jfb,Young:2022phe}). In this section, we present a comprehensive computation of the PBH mass distribution that consistently incorporates full-order non-Gaussianities. We first detail the non-linear relation between the primordial curvature perturbation and the density contrast, properly accounting for the highly non-Gaussian tail of the probability distribution function induced by the axion curvaton dynamics. Subsequently, we analyze the isocurvature perturbations carried by the PBH dark matter and discuss the stringent observational constraints they impose on the parameter space.

\subsection{The NG PBH mass distribution at all orders}

A consistent computation of the PBH abundance requires the inclusion of two distinct sources of NG. First, even if the primordial curvature perturbation $\zeta$ is exactly Gaussian, the relation between $\zeta$ and the density contrast $\delta$ is intrinsically non-linear \cite{DeLuca:2019qsy,Young:2019yug}. 
In the long-wavelength approximation, this relation reads \cite{Harada:2015yda}
\begin{equation}
\delta(\vec{x},t) = -\frac{2}{3}\Xi 
\left(\frac{1}{aH}\right)^2 e^{-2\zeta(\vec{x})} 
\Bigl[ \nabla^2\zeta(\vec{x}) + \frac12 \partial_i\zeta(\vec{x}) \, 
\partial_i\zeta(\vec{x}) \Bigr],
\label{eq:deltanl}
\end{equation}
where $a(t)$ is the scale factor, $H(t)$ the Hubble rate, and $\Xi$ is a coefficient that depends on the equation of state parameter $w = p/\rho$. For a constant $w$, one finds $\Xi = 3(1+w)/(5+3w)$; during radiation domination ($w=1/3$), we have $\Xi = 2/3$. The exponential factor and the quadratic gradient term in Eq.~\eqref{eq:deltanl} introduce unavoidable non-linearities (often referred to as \emph{non-linear NG}) even for a Gaussian $\zeta$.

Second, the primordial curvature perturbation $\zeta$ itself may exhibit intrinsic non-Gaussianity. In many inflationary scenarios, such as ultra-slow-roll inflation \cite{Atal:2019cdz} or multi-field models (e.g., the curvaton \cite{Sasaki:2006kq}), the full non-linear relation between $\zeta$ and an underlying Gaussian field $\zeta_{\rm G}$ can be resumed into a closed functional form
\begin{equation}
\zeta(\vec{x}) = \mathcal{F}\bigl(\zeta_{\rm G}(\vec{x})\bigr),
\label{eq:zetaNG}
\end{equation}
where $\mathcal{F}$ is a non-linear function. In the curvaton model with a quadratic potential, one obtains the expression given in \eqref{eq:zeta_full_nonlinear}. These \emph{primordial NGs} can significantly alter the tail of the probability distribution and thus the PBH production rate.

The PBH abundance can be computed by using the \emph{compaction function} $\mathcal{C}(r)$ \cite{Shibata:1999zs}, which measures the local mass excess within a sphere of areal radius $R(r,t) = a(t) r e^{\zeta(r)}$. It is defined as twice the mass excess divided by the areal radius \cite{Young:2022phe}: 
\begin{equation}
\mathcal{C}(r,t) \equiv \frac{2\bigl[M(r,t)-M_b(r,t)\bigr]}{R(r,t)} 
= \frac{2}{R(r,t)} \int_{S_R^2} d^3\vec{x} \; \rho_b(t) \, 
\delta(\vec{x},t),
\label{eq:Cdef}
\end{equation}
where $M(r,t)$ is the Misner–Sharp mass within the sphere, $M_b(r,t) = \rho_b(t) (4\pi/3) R^3$ is the background mass in the same volume, and $\delta = \delta\rho/\rho_b$.

The fractional mass density of the universe that collapses into PBHs at a given horizon mass $M_H$ is given by the integral over the region where the compaction exceeds the threshold $\mathcal{C}_{\rm th}$(which corresponds to the density threshold $\delta_c$ via $\mathcal{C}_{\rm th} = \delta_c$ for the volume-averaged density contrast). Including the critical collapse scaling, the mass fraction $\beta_{\rm NG}$ reads
\begin{equation}
\beta_{\rm NG}(M_H) = 
\int_{\mathcal{D}} \mathcal{K} \, 
\bigl[\mathcal{C}(\mathcal{C}_{\rm G},\zeta_{\rm G}) - 
\mathcal{C}_{\rm th}\bigr]^\gamma \, 
{\rm P}_{\rm G}(\mathcal{C}_{\rm G},\zeta_{\rm G}) \, 
d\mathcal{C}_{\rm G} \, d\zeta_{\rm G},
\label{eq:betaNG}
\end{equation}
where ${\rm P}_{\rm G}(\mathcal{C}_{\rm G},\zeta_{\rm G})$ is the two-dimensional joint probability distribution function given in the Appendix \ref{app:PBH}, and $\mathcal{C}_{\rm G}$ is the gaussian part of $\mathcal{C}$. The domain $\mathcal{D}$ is defined by the conditions
\begin{equation}
\mathcal{D} = \bigl\{ \mathcal{C}(\mathcal{C}_{\rm G},\zeta_{\rm G}) > 
\mathcal{C}_{\rm th} \;\; \wedge \;\; 
\mathcal{C}_1(\mathcal{C}_{\rm G},\zeta_{\rm G}) < 2\Xi \bigr\}.
\end{equation}
The second condition selects the type-I perturbations\cite{Musco:2020jjb} (those with a single maximum of the compaction, $\mathcal{C}_1$ being the linear component). The critical collapse parameters are numerically determined; during radiation domination one typically has $\gamma \simeq 0.36$ and $\mathcal{K} = \mathcal{O}(1\text{--}10)$\cite{Young:2019yug}, with precise values depending on the power spectrum shape\cite{Musco:2023dak}.

The PBH abundance $f_{\text{PBH}}$ can be derived from the mass fraction $\beta_{\rm NG}$ as
\begin{equation}
f_{\text{PBH}}(M_{\text{PBH}}) \equiv \frac{1}{\Omega_{\text{DM}}} \frac{d\Omega_{\text{PBH}}}{d\log M_{\text{PBH}}}, \quad \text{with} \quad \Omega_{\text{PBH}} = \int d\log M_H \left( \frac{M_{\text{eq}}}{M_H} \right)^{1/2} \beta_{\text{NG}}(M_H),
\label{eq:fPBH}
\end{equation}
where $\Omega_{\mathrm{DM}} \simeq 0.264$ is the observed cold dark matter density of the universe. $M_{\text{eq}} \approx 2.8 \times 10^{17} \, M_{\odot}$ is the horizon mass at the time of matter-radiation equality with $M_{\odot}$ being solar mass.
The relation between $M_H$ and scale $k$ and temperature $T$ is given by\cite{Franciolini:2022tfm}
\begin{equation}M_H\simeq17M_\odot\left(\frac{g_\rho}{10.75}\right)^{-1/6}\left(\frac{k/\kappa}{10^6\mathrm{~Mpc}^{-1}}\right)^{-2}=1.5\times10^{-1}M_{\odot}\left(\frac{g_{\rho}}{10.75}\right)^{-1/2}\left(\frac{T}{\mathrm{GeV}}\right)^{-2},\end{equation}
where $\kappa$ is defined in Eq.~(\ref{eq:rmk}) and $g_{\rho}$ denotes the effective numbers of relativistic degrees of freedom for energy density. The mass of the PBH formed  can be related to the horizon mass $M_H$ using
\begin{equation}
    M_{\text{PBH}} = \mathcal{K} M_H \left[ \left( \mathcal{C} - \frac{1}{4\Xi} \mathcal{C}^2 \right) - \mathcal{C}_{\text{th}} \right]^\gamma.
\end{equation}
The explicit expression of $f_{\mathrm{PBH}}$ in terms of $M_{\text{PBH}}$ is therefore
\begin{equation}
\begin{aligned}
f_{\text{PBH}}(M_{\text{PBH}}) 
&= \frac{1}{\Omega_{\text{DM}}} \int_{M_H(M_{\text{PBH}})} d\log M_H 
\left( \frac{M_{\text{eq}}}{M_H} \right)^{1/2} 
\left[ 1 - \frac{C_{\text{th}}}{\Xi} - \frac{1}{\Xi} \left( \frac{M_{\text{PBH}}}{\mathcal{K} M_H} \right)^{1/\gamma} \right]^{-1/2} \\
&\quad \times \frac{\mathcal{K}}{\gamma} \left( \frac{M_{\text{PBH}}}{\mathcal{K} M_H} \right)^{\frac{1+\gamma}{\gamma}} 
\int d\zeta_{\rm G} {\rm P}_{\rm G}(\mathcal{C}_{\rm G}(M_{\text{PBH}}, \zeta_{\rm G}), \zeta_{\rm G} | M_H) \left| \frac{d\mathcal{F}}{d\zeta_{\rm G}} \right|^{-1}.
\end{aligned}
\end{equation}

\begin{figure}[tbp]
    \centering
    \includegraphics[width=1\linewidth]{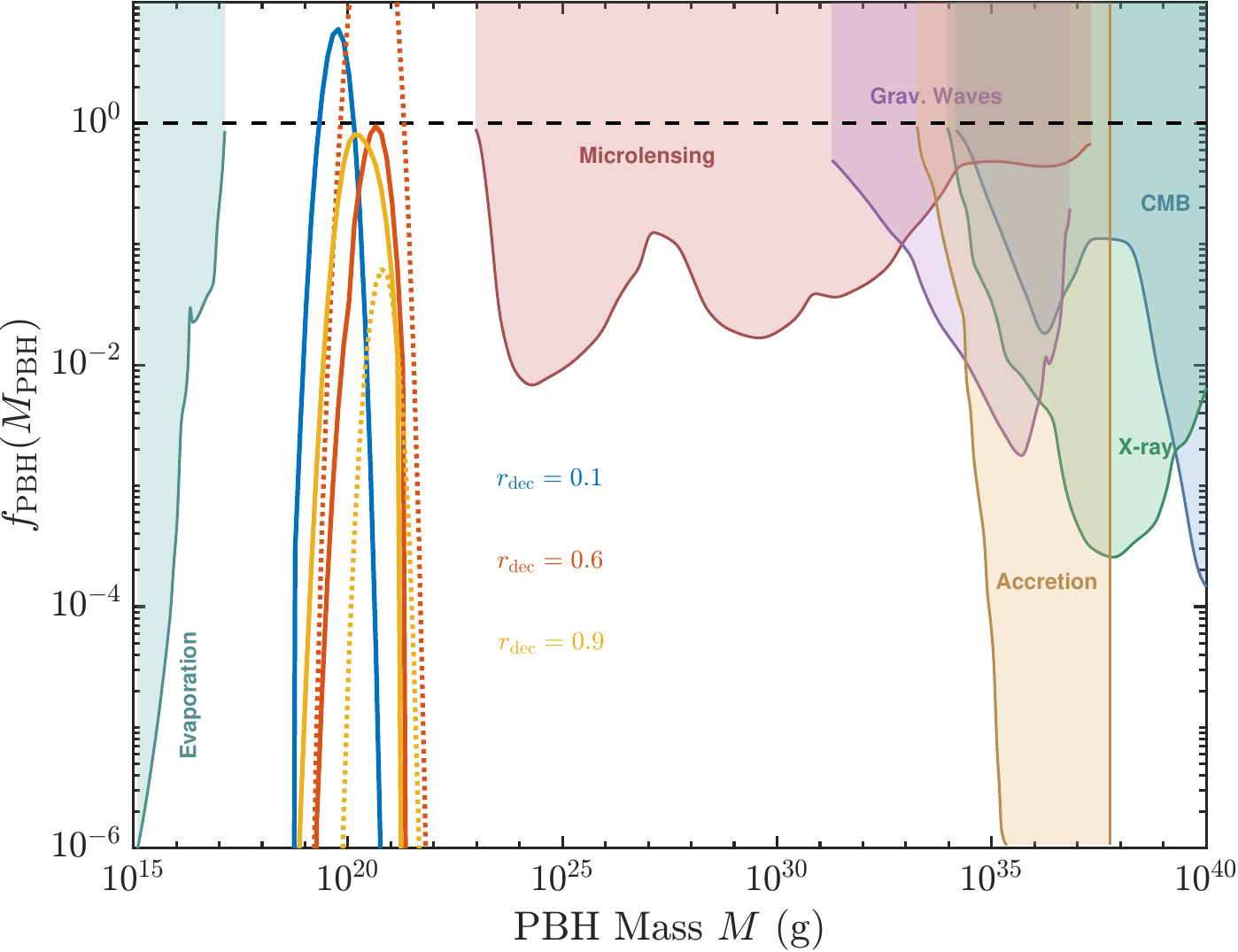}
    \caption{ 
        The fractional PBH abundance $f_{\mathrm{PBH}}$ illustrating the effects of non-Gaussianity. Dotted lines denote the baseline calculations assuming purely Gaussian curvature perturbations, while solid lines represent the results incorporating the corrections from non-Gaussianity. The colors indicate different curvaton decay fractions: $r_{\mathrm{dec}}=0.1$ (blue), $0.6$ (red), and $0.9$ (yellow). For $r_{\mathrm{dec}}=0.1$, the non-Gaussianity $f_{\mathrm{NL}}$ is positive, leading to a massive enhancement in PBH production (the corresponding dotted Gaussian line is negligibly small and falls below the plotted range). In comparison, for $r_{\mathrm{dec}}=0.6$,  $f_{\mathrm{NL}}$ is negative and it significantly suppresses the abundance. Interestingly, for $r_{\mathrm{dec}}=0.9$, although $f_{\mathrm{NL}}$ is also negative, the solid line appears higher than the dotted line. This specific behavior arises from $\mathcal{O}(1)$ theoretical uncertainties inherent in applying the current threshold-calculation framework to broad curvature power spectra. %The shaded experimental constraints match those utilized in our main results.
    }
    \label{fig:PBHgNG}
\end{figure}

The formalism described above provides a comprehensive framework for computing the PBH abundance in the presence of arbitrary local primordial non-Gaussianity, captured by the function $\mathcal{F}(\zeta_{\rm G})$. It consistently incorporates the non-linearities inherent in the relation between curvature and density perturbations. To explicitly quantify the  impact of this full-order non-Gaussianity on the PBH mass spectrum, we compute $f_{\mathrm{PBH}}$ across different curvaton decay fractions. As illustrated in Fig.~\ref{fig:PBHgNG}, the inclusion of non-linear corrections drastically alters the predicted PBH abundance compared to the naive Gaussian assumption, either exponentially enhancing or severely suppressing the formation rate depending on $r_{\mathrm{dec}}$.

However, applying this exact framework to the full axion cosine potential requires caution. In the non-perturbative regime where the initial misalignment angle is large ($\theta \sim 1$), the higher-order terms in the expansion of the cosine potential become unsuppressed and can no longer be safely truncated at the quadratic level (i.e., the harmonic oscillator approximation breaks down). This intrinsically non-perturbative dynamics causes the total curvature perturbation $\zeta$ to exhibit extreme, non-linear deviations from its underlying Gaussian component $\zeta_{\rm G}$. In particular, because of the non-monotonicity of the potential in our setup, the nature of perturbation qualitatively deviates from that of gaussian perturbations. In this regime, the standard calculation fails to provide robust predictions for the correct PBH abundance, because the critical threshold for PBH formation is fundamentally derived under the assumption of a nearly Gaussian curvature perturbation \cite{Musco:2020jjb}. Consequently, throughout this work, we restrict our analysis to the perturbative regime (as reflected in our chosen parameter sets where $\theta \lesssim 0.3$) where the quadratic approximation remains safely valid, deferring the exact treatment of the full cosine potential to future studies (see Appendix \ref{app:PBH} for further discussion).

\subsection{PBH Isocurvature Perturbations}
\label{subsec:PBH_isocurvature}

In addition to adiabatic perturbations, PBHs can also carry isocurvature perturbations if their formation is associated with a field that possesses fluctuations in an isocurvature direction. In the context of curvaton-like models, the curvaton field $\Phi$ can be decomposed into a radial mode $\varphi$ and an angular mode $\theta$. Fluctuations in the angular mode, $\delta\theta$, can induce fluctuations in the PBH abundance. Since the PBH abundance $\beta$ is exponentially sensitive to the underlying parameters, even a small modulation in $\delta\theta$ can lead to large isocurvature perturbations.

\subsubsection{Isocurvature Bounds and Non-Gaussianity}

The gauge-invariant PBH isocurvature perturbation is defined as $\mathcal{S}_{\mathrm{PBH}} = \frac{\delta\rho_{\mathrm{PBH}}}{\rho_{\mathrm{PBH}}} - \frac{3}{4}\frac{\delta\rho_r}{\rho_r}$. Evaluated on the uniform radiation density slice (where $\delta\rho_r = 0$), the local relative fluctuation in the PBH energy density is entirely determined by the spatial modulation of the PBH formation fraction $\beta$. Here, $\beta$ is exactly the fractional mass density introduced in Section~\ref{sec:PBH} (denoted there as $\beta_{\mathrm{NG}}$ to emphasize the inclusion of non-Gaussian corrections), representing the fraction of the universe's energy density that collapses into PBHs at the time of horizon reentry. Thus, the PBH isocurvature perturbation simplifies to:
\begin{equation}
\mathcal{S}_{\mathrm{PBH}} \equiv \left. \frac{\delta\rho_{\mathrm{PBH}}}{\rho_{\mathrm{PBH}}} \right|_{\delta\rho_r = 0}
= \frac{\delta\beta}{\beta}
= \frac{\partial \ln \beta}{\partial \theta} \delta\theta .
\label{eq:PBH_isocurvature_def}
\end{equation}
Recalling from Section~\ref{sec:curvper} that the primordial isocurvature perturbation seeded by the curvaton on the spatially flat slicing is $\mathcal{S} \simeq 2\delta\theta/\theta$, we can explicitly express $\mathcal{S}_{\mathrm{PBH}}$ in terms of this initial seed:
\begin{equation}
\mathcal{S}_{\mathrm{PBH}} \simeq \frac{\theta}{2} \frac{\partial \ln \beta}{\partial \theta} \mathcal{S} .
\label{eq:PBH_isocurvature_relation}
\end{equation}
This relation transparently illustrates how the primordial seed $\mathcal{S}$ is non-linearly amplified into macroscopic PBH isocurvature perturbations. To estimate the magnitude of this amplification, we consider the simplified case where $\beta$ is evaluated assuming Gaussian perturbations. For PBHs to constitute all of dark matter, the required abundance at formation is $\beta \sim 10^{-16}$ \cite{Carr:2020gox}. In the Gaussian approximation, $\beta$ is given by
\begin{equation}
\beta \sim \exp\left( -\frac{\delta_c^2}{2\sigma^2} \right),
\label{eq:beta_gaussian}
\end{equation}
where $\delta_c \approx 0.56$ is the threshold density contrast and $\sigma$ is the standard deviation of the Gaussian density perturbation $\delta_{\rm G}$. The exponent $\delta_c^2/(2\sigma^2)$ is typically $\mathcal{O}(100)$. Using the linear relation between the density contrast and the curvature perturbation at horizon crossing, $\delta_{\rm G} = (4/9)\zeta_{\rm G}$, we can express $\sigma$ in terms of the power spectrum of $\zeta_{\rm G}$.

Since the primary adiabatic curvature perturbations at large scales are generated by the inflaton, while the isocurvature perturbations originate from the independent quantum fluctuations of the curvaton's angular mode ($\delta\theta$), the resulting dark matter isocurvature is statistically \emph{uncorrelated} with the CMB adiabatic perturbations. Therefore, the uncorrelated isocurvature perturbation becomes
\begin{equation}
\mathcal{S}_{\mathrm{PBH}} \sim -\frac{\delta_c^2}{\sigma^2} \frac{\delta\theta}{\theta}.
\label{eq:S_PBH_estimate}
\end{equation}
Taking into account the dependence of $\theta$ on $r_{\mathrm{dec}}$ (see  Eq.~\eqref{eq:rD}), one can also perform numerical calculation for $\mathcal{S}_{\mathrm{PBH}}$, which is of same order as Eq.~\eqref{eq:S_PBH_estimate}. Note that the isocurvature perturbations originate from the coupling between long- and short-wave length modes, i.e., non-gaussianity of the perturbations produced by the axion curvaton. PBH clustering on long-wave length modes is computed in~\cite{Suyama:2019cst} in a model-independent way.

Current CMB constraints on isocurvature perturbations at scales $k \sim 10^{-4} - 10^{-1} \mathrm{Mpc}^{-1}$ require the isocurvature power spectrum $\mathcal{P}_{\mathrm{iso}}\equiv\mathcal{P}_{\mathcal{S}_{\mathrm{PBH}}} \lesssim 10^{-10}$ \cite{Planck:2018vyg}. For example, for $\delta\theta/\theta \sim \mathcal{O}(10^{-5})$ on large scales, the estimate in Eq.~\eqref{eq:S_PBH_estimate} would yield $\mathcal{S}_{\mathrm{PBH}} \sim \mathcal{O}(10^{-3})$, which is far above the observational bound. 

To evade this stringent bound, the amplitude of the angular field fluctuations, $\delta\theta \simeq H/(2\pi\varphi)$, must be highly suppressed when the CMB scales exit the horizon. The maximum theoretical suppression is achieved if the radial field $\varphi$ takes its maximal possible value, namely the reduced Planck mass, $\varphi_{\mathrm{max}} \sim M_{P} \approx 2.4 \times 10^{18}\,\mathrm{GeV}$. Using the relation $\mathcal{P}_{\mathcal{S}_{\mathrm{PBH}}} \simeq (\delta_c^2/\sigma^2)^2 (H / 2\pi\varphi\theta)^2$ and noting that a substantial PBH dark matter fraction requires $\delta_c^2 / (2\sigma^2) \approx 37$, we have $\delta_c^2/\sigma^2 \approx 74$. Substituting $\varphi \sim M_{P}$, a typical misalignment angle $\theta \sim \mathcal{O}(0.1)$, and the CMB constraint $\mathcal{P}_{\mathrm{iso}}^{1/2} \lesssim 10^{-5}$, we can explicitly estimate the upper bound on the inflationary Hubble scale:
\begin{equation}
    H \lesssim 2\pi \varphi_{\mathrm{max}} \theta \left( \frac{\sigma^2}{\delta_c^2} \right) \mathcal{P}_{\mathrm{iso}}^{1/2} \sim 2\pi (2.4 \times 10^{18}\,\mathrm{GeV}) (0.1) \left(\frac{1}{74}\right) (10^{-5}) \sim 2 \times 10^{11}\,\mathrm{GeV}.
\end{equation}
While $H \sim 2 \times 10^{11}\,\mathrm{GeV}$ marginally satisfies the bound at the largest observable scales ($k \sim 10^{-4}\,\mathrm{Mpc}^{-1}$), it is insufficient when the dynamical evolution of the radial field is taken into account. In our model, $\varphi$ must monotonically roll down its potential during inflation to generate the highly blue-tilted, peaked curvature spectrum required for PBH formation at much smaller scales. Consequently, as the comoving scale $k$ increases across the CMB window (from $10^{-4}$ to $10^{-1}\,\mathrm{Mpc}^{-1}$), the value of $\varphi$ decreases. Because the isocurvature power spectrum scales as $\mathcal{P}_{\mathcal{S}_{\mathrm{PBH}}} \propto \varphi^{-2}$, this dynamical evolution naturally induces a blue tilt in the isocurvature spectrum. To prevent the isocurvature perturbations from exceeding the observational upper limit at slightly smaller CMB scales (e.g., $k \sim 10^{-2}\,\mathrm{Mpc}^{-1}$), the inflationary Hubble parameter must be strictly lower than the absolute maximum derived above, forcing $H \sim 10^{10}\,\mathrm{GeV}$. At the same time, to produce a PBH abundance that accounts for all dark matter, the Gaussian curvature perturbation must satisfy $\mathcal{P}_{\zeta} \sim 0.01$, which in turn requires the symmetry-breaking scale $f$ to be of the same order as $H$, i.e., $f \sim H < 10^{10}\,\mathrm{GeV}$.

Furthermore, one might intuitively expect that the highly non-linear, exponential dependence of the PBH abundance on the field fluctuation $\delta\theta$ would generate a sizable non-Gaussianity in the isocurvature perturbations, potentially yielding a constraint stronger than that from the power spectrum itself. To rigorously assess this, we evaluate the local-type non-Gaussianity parameter for the isocurvature mode, $f_{\mathrm{NL}}^{\mathrm{iso}}$. 

In the context of CMB observations, the local-type non-Gaussianity for isocurvature perturbations is conventionally parameterized as
\begin{equation}
    \mathcal{S}_{\mathrm{PBH}} = \mathcal{S}_g + f_{\mathrm{NL}}^{\mathrm{iso}} (\mathcal{S}_g^2 - \langle \mathcal{S}_g^2 \rangle),
\end{equation}
where $\mathcal{S}_g$ is the purely Gaussian linear component. Assuming the PBH mass fraction follows the Gaussian approximation $\beta \sim \exp\left(-\frac{\delta_c^2}{2\sigma^2}\right)$, and given that the variance scales with the misalignment angle as $\sigma^2 \propto \theta^{-2}$, we can parameterize the abundance as $\beta(\theta) = \exp(-K\theta^2)$, where $K \equiv \frac{\delta_c^2}{2\sigma^2 \theta^2}$.

Expanding the isocurvature perturbation $\mathcal{S}_{\mathrm{PBH}} = \frac{\delta\beta}{\beta}$ up to second order in the field fluctuation $\delta\theta$, we obtain
\begin{equation}
    \mathcal{S}_{\mathrm{PBH}} \simeq \frac{\beta'}{\beta} \delta\theta + \frac{1}{2} \frac{\beta''}{\beta} \delta\theta^2.
\end{equation}
Identifying the linear Gaussian part as $\mathcal{S}_g = \frac{\beta'}{\beta} \delta\theta$, we can rewrite the second-order term to match the $f_{\mathrm{NL}}^{\mathrm{iso}}$ template:
\begin{equation}
    \frac{1}{2} \frac{\beta''}{\beta} \delta\theta^2 = \left[ \frac{1}{2} \frac{\beta \beta''}{(\beta')^2} \right] \mathcal{S}_g^2.
\end{equation}
This yields the exact analytical expression for the isocurvature non-linearity parameter:
\begin{equation}
    f_{\mathrm{NL}}^{\mathrm{iso}} = \frac{1}{2} \frac{\beta \beta''}{(\beta')^2}.
\end{equation}

Substituting $\beta(\theta) = \exp(-K\theta^2)$ and its derivatives $\beta' = -2K\theta \beta$ and $\beta'' = \beta(4K^2\theta^2 - 2K)$, we find
\begin{equation}
    f_{\mathrm{NL}}^{\mathrm{iso}} = \frac{1}{2} \frac{\beta \cdot \beta(4K^2\theta^2 - 2K)}{(-2K\theta \beta)^2} = \frac{1}{2} \left( 1 - \frac{1}{2K\theta^2} \right).
\end{equation}
Noting that $K\theta^2 = \frac{\delta_c^2}{2\sigma^2}$, and recalling that for PBHs to constitute a significant fraction of dark matter, the exponential suppression factor requires $\frac{\delta_c^2}{2\sigma^2} \approx 37$, we evaluate $f_{\mathrm{NL}}^{\mathrm{iso}}$ as
\begin{equation}
\label{eq:fNLDM}
    f_{\mathrm{NL}}^{\mathrm{iso}} \approx \frac{1}{2} \left( 1 - \frac{1}{2 \times 37} \right) \approx 0.49.
\end{equation}
The analytical estimate matches well with numerical calculations, as discussed in Sec.~\ref{sec:results}. Our local non-linearity parameter, $f_{\mathrm{NL}}^{\mathrm{iso}} \approx 0.49$, is normalized with respect to the isocurvature power spectrum $\mathcal{P}_{\mathcal{S}_{\mathrm{PBH}}}$. In contrast, observational limits from the Planck satellite are conventionally normalized relative to the total primordial adiabatic power spectrum $\mathcal{P}_{\Psi} = \frac{9}{25}\mathcal{P}_{\zeta}$, which is overwhelmingly dominated by the inflaton at large scales. The effective non-linearity parameter constrained by Planck, denoted as $f_{\mathrm{NL}}^{\mathrm{Planck}}$, is thus related to our intrinsic parameter by:
\begin{equation}
    f_{\mathrm{NL}}^{\mathrm{Planck}} \simeq f_{\mathrm{NL}}^{\mathrm{iso}} \left( \frac{\mathcal{P}_{\mathcal{S}_{\mathrm{PBH}}}}{\mathcal{P}_{\Psi}} \right)^2 = f_{\mathrm{NL}}^{\mathrm{iso}} \left( \frac{25 \mathcal{P}_{\mathcal{S}_{\mathrm{PBH}}}}{9 \mathcal{P}_{\zeta}} \right)^2.
\end{equation}
At the CMB pivot scale ($k = 0.05~\mathrm{Mpc}^{-1}$), the stringent observational bounds on the isocurvature fraction require $\mathcal{P}_{\mathcal{S}_{\mathrm{PBH}}} / \mathcal{P}_{\zeta} \lesssim \mathcal{O}(10^{-2})$. This suppresses $f_{\mathrm{NL}}^{\mathrm{Planck}}$ by several orders of magnitude to a value safely below $10^{-2}$. Furthermore, owing to the blue-tilted nature of the isocurvature spectrum in our model, the most rigorous test arises at smaller scales relevant to the Lyman-$\alpha$ forest ($k \sim 1~\mathrm{Mpc}^{-1}$), where $\mathcal{P}_{\mathcal{S}_{\mathrm{PBH}}}$ approaches $\mathcal{P}_{\zeta}$ (with $\mathcal{P}_{\mathcal{S}_{\mathrm{PBH}}} \lesssim 10^{-9}$ and $\mathcal{P}_{\zeta} \approx 2.1 \times 10^{-9}$). Even in this most constraining, conservatively estimated regime, the conversion factor $(25 \mathcal{P}_{\mathcal{S}_{\mathrm{PBH}}} / 9 \mathcal{P}_{\zeta})^2$ is at most $\mathcal{O}(1)$. Consequently, the effective $f_{\mathrm{NL}}^{\mathrm{Planck}}$ remains $\mathcal{O}(1)$, which is well within the current Planck bounds of $\mathcal{O}(10)$ to $\mathcal{O}(100)$ for purely isocurvature local non-Gaussianity~\cite{Planck:2019kim}. Therefore, the exponential mapping securely evades all bispectrum constraints across all relevant scales.

\subsubsection{Constraints on oscillating curvaton}

The stringent CMB bounds derived above restrict the inflationary symmetry-breaking scale $f$ to be below $ 10^{10}\,\mathrm{GeV}$.
However, such a low value of $f$ may make it impossible for the energy density of an oscillating curvaton to dominate the universe before the PBH formation epoch $t_\star$ (the time corresponding to the peak of the PBH spectrum), assuming the vacuum decay constant is $\mathcal{O}(f)$. More specifically, the energy density relation
\begin{equation}
    \frac{1}{2}m_{\sigma}^2f^2\left(\frac{H_\star}{H_{\mathrm{osc}}}\right)^{\frac{3}{2}} \gtrsim \frac{\pi^2g_{\star}}{30}T_{\star}^4
    \label{eq:mmff}
\end{equation}
may not be satisfied, which leads to insufficient PBH production (It is true that scenario with $r_D<1$ could still generate large fraction of PBH through non-Gaussian effect. However for $r_D\ll1$ the PBH production falls quickly due to small curvature perturbation). This is a direct consequence of the fact that the curvaton energy fraction at decay scales as $\rho_\sigma/\rho_{\mathrm{tot}} \propto f^2$, and a small $f$ leads to a subdominant curvaton.

To resolve this tension, we introduce a non-renormalizable superpotential that allows the curvaton vacuum expectation value (VEV) after inflation to be significantly larger than the scale $f$ appearing during inflation. Consider the superpotential
\begin{equation}
W = \frac{1}{n+1} \frac{\bar{\Phi}^{\,n+1}}{M_{P}^{\,n-2}},
\label{eq:superpotential}
\end{equation}
which generates the following scalar potential for the complex field $\bar{\Phi} = \varphi e^{i\theta}$:
\begin{equation}
V(\bar{\Phi}) = \frac{1}{M_{P}^{2n-4}} |\bar{\Phi}|^{2n}
- m_S^2 |\bar{\Phi}|^2
+ c_2 H^2 |\bar{\Phi}|^2
+ \left( A \frac{\bar{\Phi}^{\,n+1}}{M_{P}^{\,n-2}} + \mathrm{h.c.} \right).
\label{eq:full_potential}
\end{equation}
%\LTW{Could there also be a term of the form $H \bar \Phi^{n+1}/M_P^{n-2}$?} \KH{The requires large $R$ symmetry breaking during inflation that is not so common for inflation models. Roughly speaking, this is because $F^2 =( \partial W/\partial \phi)^2 \sim W^2/\phi^2$, so unless $\phi \sim Mpl$, $W^2/Mpl^2$ is negligible. Even if $\phi \sim Mpl$, models with $F \sim W/Mpl$ has an instability problem from $V =F^2 - 3 W^2 Mpl^2$ becoming negative.}
Here $m_S^2 > 0$ is a soft symmetry-breaking mass term, $c_2$ is a dimensionless coupling, and $A$ is a complex parameter of order $m_S$ that explicitly breaks the global $U(1)$ symmetry. Since $c_2$ is assumed to be negative, $\bar{\Phi}$ is driven toward large values during inflation. After inflation, as the Hubble parameter $H$ decreases, $\bar{\Phi}$ adiabatically tracks the minimum determined by the balance between the first and third terms~\cite{Dine:1995kz,Harigaya:2015hha}. 
When $H$ drops below the soft breaking scale $m_S$, the radial field is settled to the true minimum, which is determined by the balance between the first and second term. This yields a post-inflationary VEV
\begin{equation}
F \equiv \langle \varphi \rangle_{\mathrm{after}} 
\sim \left( M_{P}^{n-2} \, m_S \right)^{1/(n-1)},
\label{eq:f_a}
\end{equation}
which can be much larger than $f$. 
The explicit breaking term (proportional to $A$) gives a mass to the curvaton (angular) mode, $m_\sigma^2 \sim A m_S$. 
Because of $F\gg f$, the curvaton energy density during the oscillatory phase, whose initial value is $\rho_\sigma \sim m_\sigma^2 F^2$, can easily dominate the universe even if $f$ is constrained to be low, provided that $F$ is sufficiently high.

Remarkably, in this setup, the field perturbation of the axion field becomes much above $H_{\rm inf}/(2\pi)$ because of $F\gg f$. This allows for large $F$, which is necessary for the axion curvaton to dominates the universe, and small enough $H_{\rm inf}$, which is necessary to avoid the constraints from dark-matter isocurvature perturbations, while generating large curvature perturbations $\zeta$.

To ensure that the curvaton decays before the PBH-forming scales re-enter the horizon, we require $T_{\mathrm{dec}} \gg T_*$, where $T_*$ is the temperature at PBH formation. Assuming the curvaton decays into radiation through an effective interaction suppressed by the vacuum decay constant $F$ (similar to an axion coupling to gauge bosons via anomaly terms), its decay rate can be parameterized as
\begin{equation}
\Gamma_\sigma = \frac{\alpha^2}{16\pi} \frac{m_\sigma^3}{F^2}, \label{eq:decay_rate}
\end{equation}
where $\alpha$ is a dimensionless effective coupling constant. We expect $\alpha \lesssim 1$; for example, if $\bar{\Phi}$ has a Yukawa coupling $y$ to other fields, the decay rate is $y^2 m_\sigma / (16\pi)$. The mass of these other fields, $y F$, must be below $m_\sigma$, capping the decay rate at a maximum of $m_\sigma^3/(16\pi F)$.
The decay temperature of the curvaton is then given by
\begin{equation}
T_{\mathrm{dec}} = \left( \frac{90}{\pi^2} \right)^{1/4} g_*^{-1/4} (\Gamma_\sigma M_{P})^{1/2}
= \frac{90^{1/4}}{4\pi} \, g_*^{-1/4} \, \alpha \, \frac{M_{P}^{1/2} m_\sigma^{3/2}}{F}.
\label{eq:T_dec}
\end{equation}

\begin{figure}[tbp]
    \centering
    \includegraphics[width=0.8\linewidth]{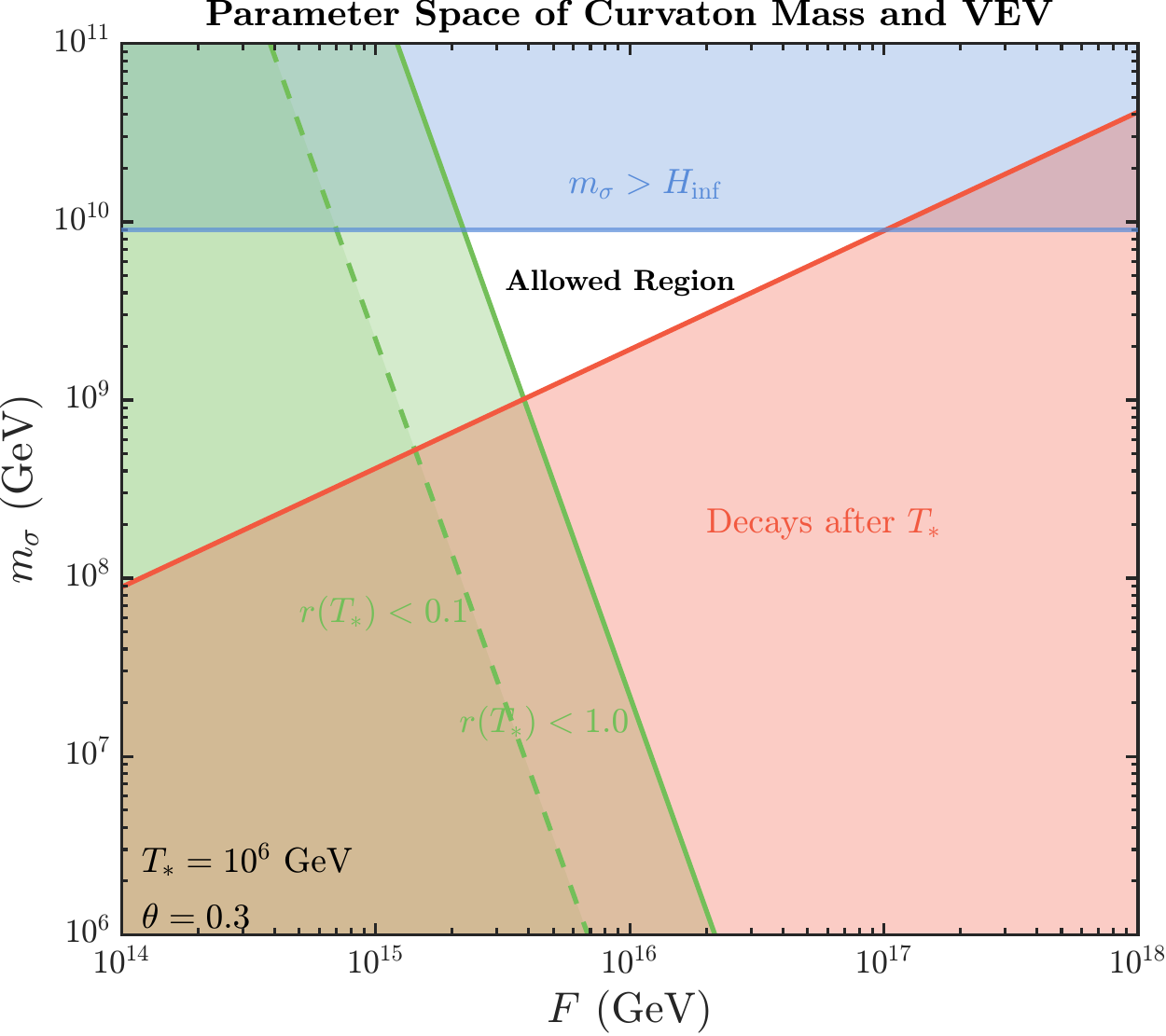}
   \caption{ Parameter space of the curvaton mass $m_\sigma$ versus the post-inflationary vacuum expectation value $F$ for an oscillating curvaton. The shaded areas denote the parameter space that is excluded or inconsistent with our assumption of radiation domination at $T=T_*$, while the unshaded region represents the allowed parameter space viable for primordial black hole (PBH) formation. The green contours delineate the lower bounds required for the curvaton to achieve a sufficient energy fraction $r$ at the PBH formation epoch ($T_*$). The solid green line marks the exact threshold for absolute energy domination ($r(T_*) = 1.0$), whereas the dashed green line marks the threshold for partial domination ($r(T_*) = 0.1$), which remains sufficient for PBH production amplified by non-Gaussianity. Parameter regions below these respective lines (green shading) fail to reach the target energy fractions and are thus excluded.
   Below the red line, the curvaton cannot decay at $T>T_*$ (derived from Eq.~(\ref{eq:T_dec})), which is inconsistent with our computation that assumes radiation domination at $T=T_*$
   Finally, the horizontal solid blue line represents the theoretical inflationary upper limit ($m_\sigma = H_{\mathrm{inf}}$), above which the parameter space is excluded. The fixed parameters are $T_* = 10^6\mathrm{~GeV}$, $H_{\mathrm{inf}} = 9 \times 10^9\mathrm{~GeV}$, and $\theta = 0.3$, targeting the data in Sec.~\ref{sec:results}.}
    \label{fig:msigma_F_space}
\end{figure}

Fig.~\ref{fig:msigma_F_space} shows the constraints on $F$ and $m_\sigma$ for $T_* \simeq 10^6 \,\mathrm{GeV}$, which corresponds to the scenario where PBHs constitute the entirety of dark matter.
Below the red line, the curvaton decays at $T_{\rm dec}<T_*$, which is inconsistent with our assumption that the universe is radiation dominated at $T=T_*$. Here we assume the maximal possible decay rate, $\alpha = 1$.
The green lines are the contours of the curvaton energy fraction $r$ at $T_*$ when the curvaton decay at $T=T_*$, which is the maximal possible energy fraction.
$r(T_*)$ should be at least $O(0.1)$ to produce PBH dark matter. In the gray-shaded region, $m_\sigma >H_{\rm inf}$. Because $m_\sigma \sim \sqrt{Am_S} \lesssim m_S$,  the dynamics of $\phi$ is no longer determined by the Hubble-induced mass in the gray-shaded region.
One can see that in the viable parameter region $m_\sigma > 10^{8}$ GeV, which requires a large supersymmetry breaking scale.
For heavier PBHs, since $T_*$ is smaller, the viable parameter space is wider.

In summary, the introduction of a higher-dimensional superpotential allows a separation between $f$ and the post-inflationary VEV $F$. This mechanism successfully decouples the isocurvature constraint (which forces $f$ to be small) from the requirement of curvaton energy domination (which benefits from a large $F$), thereby opening a viable parameter space for PBH dark matter in curvaton models.

Alternatively, one can suppress the angular fluctuations by directly coupling the curvaton to the inflaton. For example, a potential of the form
\begin{equation}
c_I H^2 \frac{\Phi^n}{M_{P}^{n-2}}
\label{eq:inflaton_coupling}
\end{equation}
can provide an additional mass of order \(H\) to the angular direction during inflation. By fine-tuning \(c_I\), one can not only cancel the Hubble-induced mass term for the radial mode, thereby keeping \(\phi\) large and slowly rolling across CMB scales, but also simultaneously endow the angular mode \(\theta\) with an effective mass \(m_\theta \gtrsim H\). This direct mass term for the angular direction significantly suppresses the quantum fluctuations \(\delta\theta\) beyond the naive \(\delta\theta \sim H/(2\pi\phi)\) estimate, because the fluctuation amplitude is reduced by a power-law depending on \(m_\theta/H\). As a result, the isocurvature perturbations sourced by \(\delta\theta\) are drastically weakened.
By numerically solving the evolution of the fields, we find that the energy scales \(f\) and \(H\) may be as high as \(10^{13}\,\mathrm{GeV}\), albeit at the cost of requiring precise tuning of \(c_I\). 

\subsubsection{Constraints on rotating curvaton}

The constraints on a rotating axion curvaton are significantly weaker. After the initiation of the rotation, the rotating axion eventually dominates the universe and reheats it through the washout of the approximately conserved $U(1)$ charge associated with the rotation~\cite{Co:2022qpr}. We consider the case where the washout rate is given by
\begin{equation}
    \Gamma = \alpha^2 \frac{T^3}{\phi^2},
\end{equation}
where $\alpha$ is a constant much smaller than unity. For example, if the washout requires the violation of the $U(1)$ symmetry by a Yukawa interaction $y$, then $\alpha^2 \sim y^2/(8\pi)$, while if a sphaleron process of the gauge interaction with a coupling $g$ is required, $\alpha^2\sim (g^2/4\pi)^4$~\cite{Co:2019wyp}. We assume that $\phi > f$ when the washout occurs, so that the rotation behaves as matter prior to the washout.%
\footnote{
If the rotation continues down to $\phi \simeq f$, it behaves as kination. The sudden change of the equation of state and/or the field perturbations enhanced by the kination phase can generate extra GWs~\cite{Harigaya:2023mhl,Bodas:2025wef}. The enhanced field perturbations can also produce axion dark matter~\cite{Eroncel:2025qlk,Bodas:2025eca}, so there can be a constraint from the overproduction of axion dark matter. 
}
Using $\Gamma \simeq 3 H$ and $m_\phi^2 \phi_{\mathrm{RH}}^2 = \pi^2 g_* T_{\mathrm{RH}}^4/30$, we find
\begin{align}
    T_{\mathrm{RH}} &\simeq 3 \times 10^7\,\mathrm{GeV} \left( \frac{m_\phi}{10^6\,\mathrm{GeV}}\right)^{2/3} \left(\frac{\alpha^2}{10^{-6}}\right)^{1/3}, \\
    \phi_{\mathrm{RH}} &\simeq 9 \times 10^9\,\mathrm{GeV} \left( \frac{m_\phi}{10^6\,\mathrm{GeV}}\right)^{1/3} \left(\frac{\alpha^2}{10^{-6}}\right)^{2/3}.
\end{align}

\begin{figure}[tbp]
    \centering
    \includegraphics[width=0.8\linewidth]{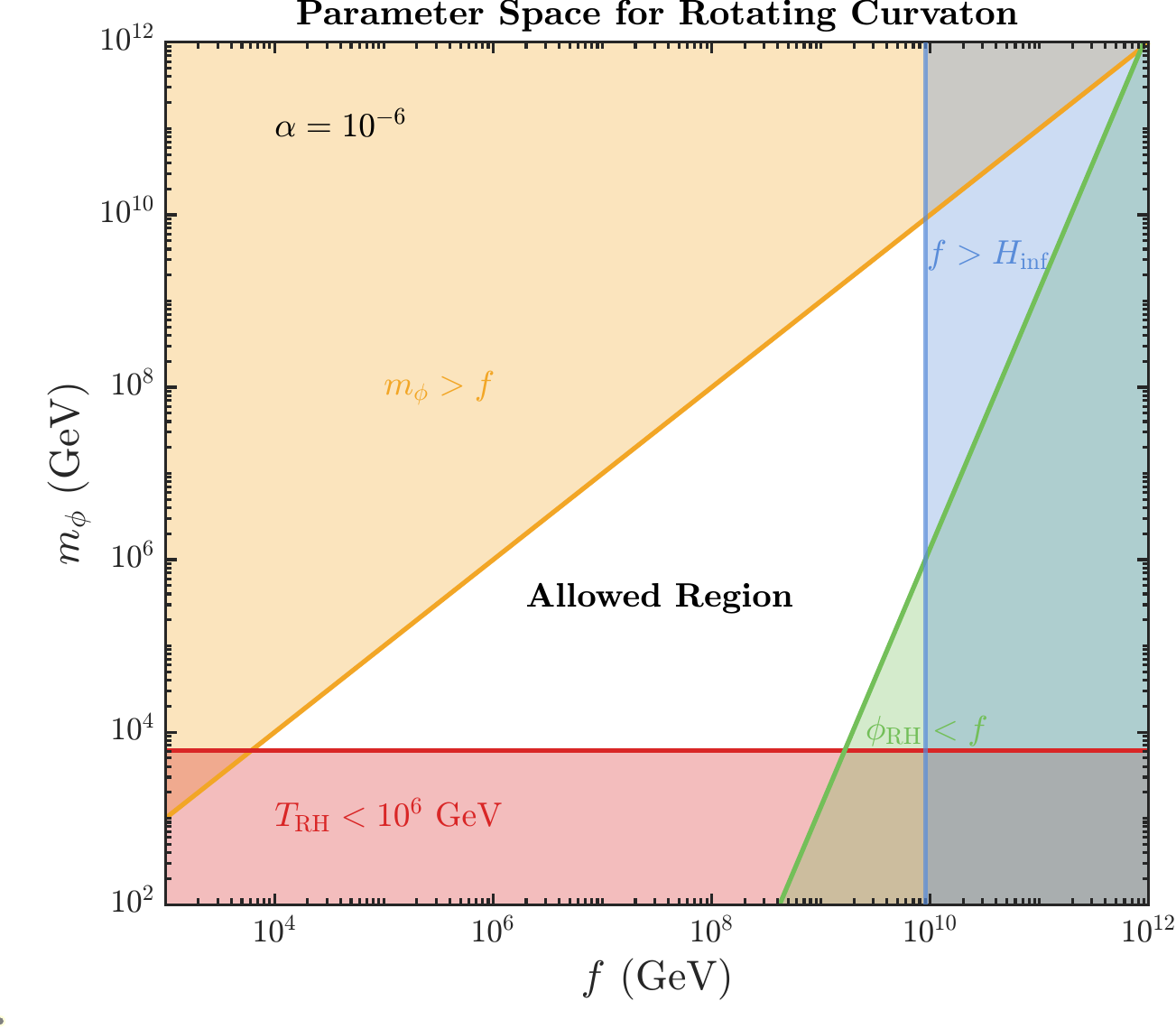}
\caption{
Parameter space of the symmetry-breaking scale $f$ versus the radial mass $m_\phi$ for a rotating curvaton. The shaded regions indicate parameter sets that are either theoretically inconsistent or incompatible with our assumption of a radiation-dominated universe at the PBH formation epoch $T_*$. The unshaded region denotes the viable parameter space for PBH production. The red shaded region corresponds to $T_{\mathrm{RH}} < T_*$, i.e., the reheating via washout completes after the PBH-forming scales re-enter the horizon, which violates our assumption of radiation domination at $T=T_*$. The green shaded region marks $\phi_{\mathrm{RH}} < f$, where the rotating field value at reheating falls below the symmetry-breaking scale; in this regime the rotation does not behave as matter-like prior to washout, invalidating the derivation of the reheating temperature and the energy fraction. The yellow shaded region indicates $m_\phi > f$, where the radial direction cannot remain on the flat trajectory $\Phi\bar{\Phi}=f^2$ during inflation. The blue shaded region corresponds to $f$ exceeding the maximal possible inflationary Hubble scale ($H_{\mathrm{inf}} \sim 9\times10^9$ GeV); in this case, the amplitude of the curvature perturbations at the peak is insufficient to produce PBHs as dark matter, even after accounting for non-Gaussianity. We take $T_* = 10^6$ GeV, with the washout coefficient taken conservatively as $\alpha^2 = 10^{-6}$ (larger values of $\alpha$ would further enlarge the viable region). 
%The boundary between the red and unshaded regions is determined by Eq.~(\ref{eq:TRH}), which sets $T_{\mathrm{RH}}(m_\phi,f) = T_*$. The green boundary is set by $\phi_{\mathrm{RH}}(m_\phi,f) = f$, and the yellow boundary by $m_\phi = f$. The blue boundary is set by $f = H_{\mathrm{inf}}$. 
%Unlike the oscillating curvaton case, the rotating curvaton enjoys a significantly broader allowed parameter space because the washout process can efficiently reheat the universe even when the radial field value is small, allowing easy satisfaction of both energy domination and early reheating.
}
    \label{fig:mphi_f_space}
\end{figure}

In Fig.~\ref{fig:mphi_f_space}, we show the constraint on $f$ and $m_\phi$ for $T_*=10^6$ GeV.  In the yellow-shaded region, $m_\phi >f$, for which $\Phi$ and $\bar{\Phi}$ cannot follow $\Phi \bar{\Phi} = f^2$. In the red-shaded region, the reheating occurs at $T<T_*$, which is inconsistent with our assumption that the universe is radiation dominated at $T_*$. In the green-shaded region, $\phi_{\rm RH} <f$. In the blue-shaded region, $f$ is larger than the maximal possible inflation scale $\sim 9\times 10^{9}$ GeV, and the magnitude of the perturbations at the peak is not large enough to produce PBH dark matter.
In deriving the reheating constraints, we took a conservative value $\alpha^2 \sim 10^{-6}$. Larger $\alpha$ allows for an even wider viable parameter space.
Unlike the oscillating curvaton, the reheating constraints can be easily satisfied. This is essentially because the value of $\phi$ around the washout may be much smaller than the initial value at the onset of the rotation, so that the washout may be efficient while the rotation dominates the universe.

\section{Gravitational Waves}
\label{sec:GW}

With the primordial curvature power spectrum derived in the previous section, we can now compute the corresponding stochastic background of gravitational waves induced at second order by scalar perturbations. These scalar-induced gravitational waves (SIGWs) arise from the nonlinear coupling of scalar modes during the radiation-dominated era. In the subhorizon limit, where the wavelength of the perturbations is much smaller than the Hubble radius, the energy density parameter of SIGWs per logarithmic wave-number interval is given by the convolution integral \cite{Kohri:2018awv, Espinosa:2018eve, Cai:2019cdl}:
\begin{equation}
\tilde{\Omega}_{\mathrm{GW}}(k) = \int_{0}^{\infty} \mathrm{d}\bar{v} \int_{|1-\bar{v}|}^{1+\bar{v}} \mathrm{d}\bar{u} \; \mathcal{K}(\bar{u},\bar{v}) \, \mathcal{P}_{\zeta}(\bar{u}k) \, \mathcal{P}_{\zeta}(\bar{v}k), \label{eq:OmegaGW_integral}
\end{equation}
where $\tilde{\Omega}_{\mathrm{GW}}(k)$ denotes the dimensionless energy density fraction evaluated at a time when the GW mode is deep inside the horizon during radiation domination. The integration variables $\bar{u}$ and $\bar{v}$ are dimensionless rescalings of the wave-numbers, defined as $\bar{u} \equiv |\mathbf{k} - \mathbf{k}'|/k$ and $\bar{v} \equiv k'/k$, and $\mathcal{K}(\bar{u},\bar{v})$ is a kernel function that encapsulates the transfer functions of the scalar perturbations and the Green's function for tensor mode propagation.

The kernel function $\mathcal{K}(\bar{u},\bar{v})$ can be derived analytically under the assumption of a perfect radiation-dominated background and instantaneous horizon crossing~\cite{Kohri:2018awv, Espinosa:2018eve},
\begin{align}
\mathcal{K}(\bar{u},\bar{v}) = & \frac{3\bigl(4\bar{v}^{2} - (1+\bar{v}^{2}-\bar{u}^{2})^{2}\bigr)^{2} \bigl(\bar{u}^{2}+\bar{v}^{2}-3\bigr)^{4}}{1024\,\bar{u}^{8}\bar{v}^{8}} \nonumber \\ 
&\Biggl[ \Biggl( \ln\Bigl| \frac{3-(\bar{u}+\bar{v})^{2}}{3-(\bar{u}-\bar{v})^{2}} \Bigr| - \frac{4\bar{u}\bar{v}}{\bar{u}^{2}+\bar{v}^{2}-3} \Biggr)^{2} 
+ \pi^{2} \, \Theta\bigl(\bar{u}+\bar{v}-\sqrt{3}\bigr) \Biggr], \label{eq:kernel}
\end{align}
where $\Theta(x)$ is the Heaviside step function. The logarithmic term arises from the integral over the conformal time, while the Heaviside term reflects the resonant production of gravitational waves when the combined scalar mode frequencies match the tensor mode frequency.

To connect the primordial GW energy density with the observed value today, we must account for the redshifting of gravitational radiation after matter-radiation equality. The present-day energy density parameter of SIGWs is given by \cite{Espinosa:2018eve}
\begin{equation}
\Omega_{\mathrm{GW}} h^{2} = 0.39 \left(\frac{g_{\rho}}{106.75}\right) \left(\frac{g_{s}}{106.75}\right)^{-4/3} \Omega_{r,0} h^{2} \; \tilde{\Omega}_{\mathrm{GW}}, \label{eq:OmegaGW_today}
\end{equation}
where $\Omega_{r,0} h^{2} \simeq 4.2 \times 10^{-5}$ is the current radiation density parameter (including photons and three neutrino species), $g_{\rho}$ and $g_{s}$ are the effective numbers of relativistic degrees of freedom for energy density and entropy, respectively, at the time of GW production, and $h \equiv H_{0}/(100\;\mathrm{km\,s^{-1}\,Mpc^{-1}})$ is the dimensionless Hubble constant. The numerical prefactor 0.39 accounts for the change in the effective relativistic degrees of freedom between GW production and today.

In our axion-like curvaton model, the curvature power spectrum $\mathcal{P}_{\zeta}(k)$ entering Eq.~(\ref{eq:OmegaGW_integral}) is given by Eq.~(\ref{eq:Pzeta_curv_final}). The resulting GW spectrum inherits the peaked shape of the scalar spectrum, but is further broadened by the double integration over wave-numbers. Because the kernel $\mathcal{K}(\bar{u},\bar{v})$ is sharply peaked when $\bar{u} \sim \bar{v} \sim 1$, the GW spectrum is dominated by scalar modes with wave-numbers near $k$.

\section{Results}
\label{sec:results}

In this section, we present the numerical results of our axion curvaton model. To concretely demonstrate the outcomes, the benchmark parameter sets evaluated in this section are based on the oscillating curvaton scenario. The rotating curvaton scenario features a similar curvature power spectrum shape and would thus generate consistent phenomenological signatures. We consider two distinct phenomenological scenarios to illustrate the production of PBHs and their accompanying observational signatures. Furthermore, we explore the intrinsic correlation between the GW spectrum and the large-scale dark-matter isocurvature perturbations of our model. This correlation offers a unique multi-messenger link that can be tested by future GW interferometers and CMB/Lyman-$\alpha$ observations.

\subsection{Baseline Scenarios: Asteroid and Solar Mass PBHs}

%\LTW{float the table and figures to match well with the text. }
\begin{table}[tbp]
    \centering
    \caption{Summary of the parameter sets used in the numerical analysis. \textbf{Group 1} corresponds to parameters yielding asteroid-mass PBHs capable of explaining the total dark matter abundance. \textbf{Group 2} corresponds to parameters yielding solar-mass PBHs. The parameters $H$ and $f$ are given in GeV. For all parameter sets in this scenario, we take the effective mass $c_2 = -1$ for simplicity, as the constraint on it is not severe.}
    \label{tab:pbh_parameters}
    \renewcommand{\arraystretch}{1.2} 
    \begin{tabular}{l c c c c c c}
        \toprule
        \textbf{Scenario} & $\boldsymbol{r_{\mathrm{dec}}}$ & $\boldsymbol{H~[10^{10} \text{ GeV}]}$ & $\boldsymbol{f~[10^{10} \text{ GeV}]}$ & $\boldsymbol{c_1}$ & 
        $\boldsymbol{c_2}$ &
        $\boldsymbol{\theta}$ \\
        \midrule
        \multicolumn{7}{l}{\textit{\textbf{Group 1: Total Dark Matter (Asteroid Mass)}}} \\
       Set 1 & $0.1$ & $0.9$ & $0.3$ & $1.21$ &$-1.00$ & $0.30$ \\
        Set 2 & $0.6$ & $1.0$ & $1.0$ & $1.21$ &$-1.00$ & $0.30$ \\
        Set 3 & $0.9$ & $0.9$ & $1.2$ & $1.21$ &$-1.00$ & $0.38$ \\
        \midrule
        \multicolumn{7}{l}{\textit{\textbf{Group 2: Solar Mass PBH}}} \\
         Set 4 & $0.1$ & $1.5$ & $0.6$ & $1.74$ &$-1.00$ & $0.20$ \\
        Set 5 & $0.6$ & $0.8$ & $0.8$ & $1.72$ &$-1.00$ & $0.24$ \\
        Set 6 & $0.9$ & $2.1$ & $2.7$ & $1.66$ &$-1.00$ & $0.30$ \\
        \bottomrule
    \end{tabular}
    \vspace{0.5cm}
     \caption{
    Same as Table~\ref{tab:pbh_parameters}, but $c_2$ is allowed to approach or be equal to zero, broadening the resulting spectra. 
    }
    \label{tab:pbh_parametersb}
    \renewcommand{\arraystretch}{1.2} 
    \begin{tabular}{l c c c c c c}
        \toprule
        \textbf{Scenario} & $\boldsymbol{r_{\mathrm{dec}}}$ & $\boldsymbol{H~[10^{10} \text{ GeV}]}$ & $\boldsymbol{f~[10^{10} \text{ GeV}]}$ & $\boldsymbol{c_1}$ & 
        $\boldsymbol{c_2}$ &
        $\boldsymbol{\theta}$ \\
        \midrule
        \multicolumn{7}{l}{\textit{\textbf{Group 1: Total Dark Matter (Asteroid Mass)}}} \\
       Set 1 & $0.1$ & $0.9$ & $0.3$ & $1.21$ &$0.00$ & $0.31$ \\
        Set 2 & $0.6$ & $1.0$ & $1.0$ & $1.21$ &$0.00$ & $0.31$ \\
        Set 3 & $0.9$ & $0.9$ & $1.2$ & $1.21$ &$-0.16$ & $0.41$ \\
        \midrule
        \multicolumn{7}{l}{\textit{\textbf{Group 2: Solar Mass PBH}}} \\
         Set 4 & $0.1$ & $2.1$ & $0.6$ & $1.74$ &$0.00$ & $0.29$ \\
        Set 5 & $0.6$ & $1.0$ & $0.8$ & $1.72$ &$0.00$ & $0.31$ \\
        Set 6 & $0.9$ & $2.0$ & $2.7$ & $1.66$ &$-0.16$ & $0.31$ \\
        \bottomrule
    \end{tabular}
\end{table}
The first scenario (Group 1) is designed to form asteroid-mass PBHs ($M \sim 10^{17}-10^{20}$ g), which can constitute the totality of dark matter ($f_{\mathrm{PBH}} \approx 1$). The second scenario (Group 2) targets the formation of solar-mass PBHs ($M \sim M_{\odot}$), which is  relevant for the gravitational wave signals observed by LIGO/Virgo. 

The specific parameter choices for the Hubble rate during inflation $H$, the decay constant $f$, the effective mass parameters $c_{1,2}$, the initial misalignment angle $\theta$, and the decay parameter $r_{\mathrm{dec}}$ are summarized in Table \ref{tab:pbh_parameters}. Here $c_1$, which determines the IR slope of the spectrum, is chosen so that the PBH dark-matter isocurvature perturbation constraint is saturated. The parameter $c_2$ is not subject to such constraints. For all baseline parameter sets in this subsection, we set $c_2$ equal to $-1$ for simplicity. For the solar-mass scenario (Group 2), we adopt a larger $c_1$ compared to the asteroid-mass scenario to adjust the IR slope.

To demonstrate the versatility of the axion curvaton framework, we also consider an alternative phenomenological regime by relaxing the constraints on the Hubble-induced mass parameter $c_2$, presented in Table \ref{tab:pbh_parametersb}. Specifically, we find that $c_2$ can be relaxed to zero while keeping PBH evaporation safely within bounds. 

It is worth noting why the curvature power spectrum still develops a peak even when $c_2 \approx 0$. In this regime, the radial field $\varphi$ continues to evolve slowly toward smaller values, driven steadily by the positive Hubble-induced mass $c_1$. Because it lacks the rapid acceleration previously provided by the negative $c_2$ term, $\varphi$ approaches the symmetry-breaking scale $f$ much more gradually before its evolution eventually halts. This prolonged slow-roll dynamics naturally broadens the spectral feature, generating a significantly wider PBH mass distribution and, consequently, a broad plateau in the GW spectrum.

Before proceeding to the detailed results, a brief remark on the cosmic thermal history is in order. For the solar-mass PBH scenarios (Group 2) and their accompanying nano-Hz GW signals, the relevant comoving scales re-enter the Hubble horizon at temperatures close to the QCD phase transition ($T \sim 100 \text{ MeV}$). It is well known that the softening of the equation of state and the rapid variation of the relativistic degrees of freedom during this epoch can induce a secondary enhancement in the PBH mass function and introduce $\mathcal{O}(1)$ modulations in the spectral tilt of the stochastic gravitational wave background (see e.g.\cite{Franciolini:2022tfm,Escriva:2022bwe,Musco:2023dak}). However, the primary focus of our work is the asteroid-mass PBH dark matter (Group 1), which forms deep within the radiation-dominated era ($T \sim 10^6 \text{ GeV}$), rendering QCD effects entirely irrelevant. For the solar-mass counterpart, we retain the approximation of a constant radiation equation of state ($w=1/3$). This treatment is sufficient to demonstrate the versatility of our model in generating the required broad or peaked spectra, as the precise incorporation of the QCD crossover would only provide sub-leading $\mathcal{O}(1)$ refinements without altering the fundamental phenomenological conclusions (see \cite{Abe:2020sqb,Ferrante:2023bgz} for this effect).

\begin{figure}[t!]
    \centering
    \begin{subfigure}[b]{0.8\textwidth}
        \centering
        \includegraphics[width=\textwidth]{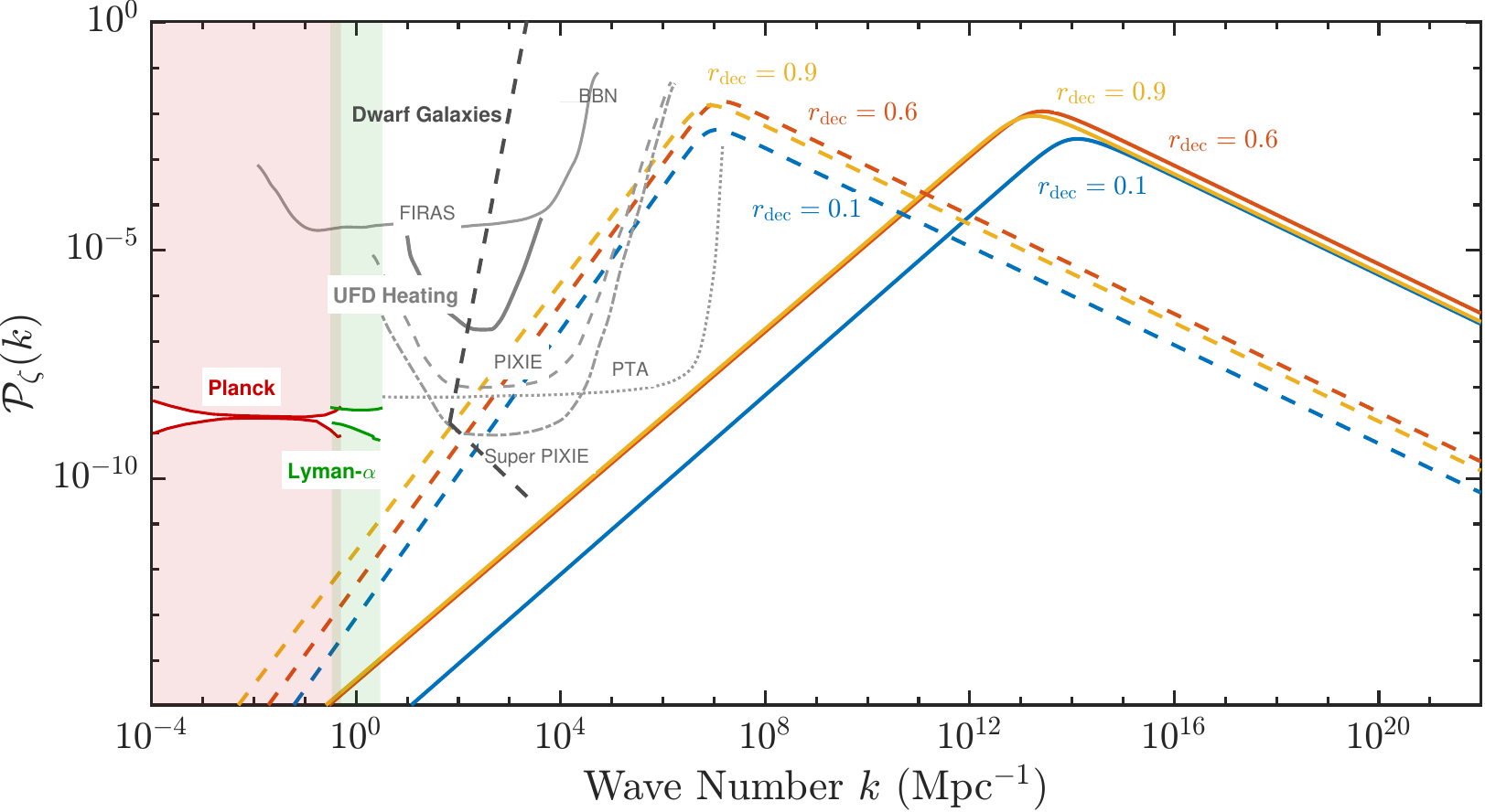}  
        \caption{Peaked primordial curvature power spectrum}
    \end{subfigure}
    \hfill
    \begin{subfigure}[b]{0.8\textwidth}
        \centering
        \includegraphics[width=\textwidth]{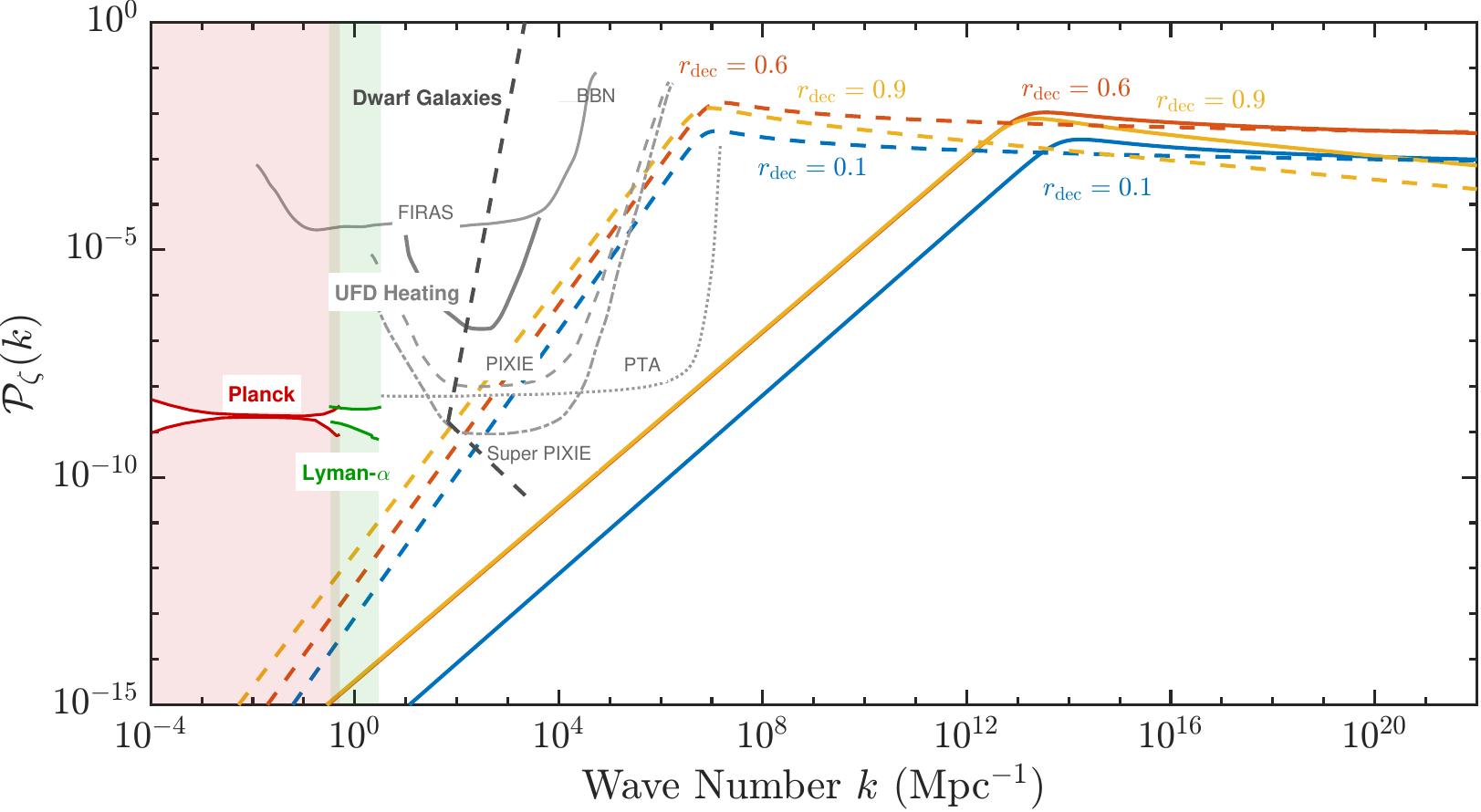} 
        \caption{Broad primordial curvature power spectrum}
    \end{subfigure}   
    \caption{The primordial curvature power spectrum $\mathcal{P}_{\zeta}(k)$ as a function of the comoving wavenumber $k$ for the parameter sets in Table a) \ref{tab:pbh_parameters} and b) \ref{tab:pbh_parametersb}. The theoretical predictions are divided into Group 1 (solid lines, peaked at $k \sim 10^{13}$ Mpc$^{-1}$ for asteroid-mass PBHs) and Group 2 (dashed lines, peaked at $k \sim 10^{7}$ Mpc$^{-1}$ for solar-mass PBHs). Within each group, the blue, red, and yellow curves denote the decay parameter values $r_{\mathrm{dec}}=0.1, 0.6$, and $0.9$, respectively. The shaded regions illustrate current observational upper bounds from Planck\cite{Planck:2018jri} and Lyman-$\alpha$\cite{Bird:2010mp,Chabanier:2019eai}. We show constraints from FIRAS, and projected reaches of PIXIE and Super-PIXIE\cite{Fixsen:1996nj,Mather:1993ij,Chluba:2012gq,Chluba:2012we,Cyr:2023pgw}, along with direct PTA observations\cite{Lee:2020wfn}, ultra-faint dwarf galaxies\cite{Graham:2024hah}, and dwarf galaxies\cite{Dekker:2024nkb}.
} 
    \label{fig:PSlisa}
\end{figure}

Utilizing the parameters summarized in Table~\ref{tab:pbh_parameters}, we compute the primordial curvature power spectrum $\mathcal{P}_{\zeta}(k)$, shown in the upper panel of Fig.~\ref{fig:PSlisa}. The spectra exhibit a distinct peak whose position is determined by the specific group parameters. Group 1 spectra peak at  small scales ($k \sim 10^{13}$ Mpc$^{-1}$), whereas Group 2 spectra peak at intermediate scales ($k \sim 10^{7}$ Mpc$^{-1}$). The choices of decay parameter ($r_{\mathrm{dec}}=0.1, 0.6, 0.9$) modulate the spectral shape and amplitude. The resulting curvature power spectrum based on Table~\ref{tab:pbh_parametersb} is presented in the lower panel of Fig.~\ref{fig:PSlisa}. All generated spectra comfortably evade current observational bounds on large scales, such as those from Planck and FIRAS.

%\clearpage
\begin{figure}[t!]
    \centering
    \begin{subfigure}[b]{0.7\textwidth}
        \centering
        \includegraphics[width=\textwidth]{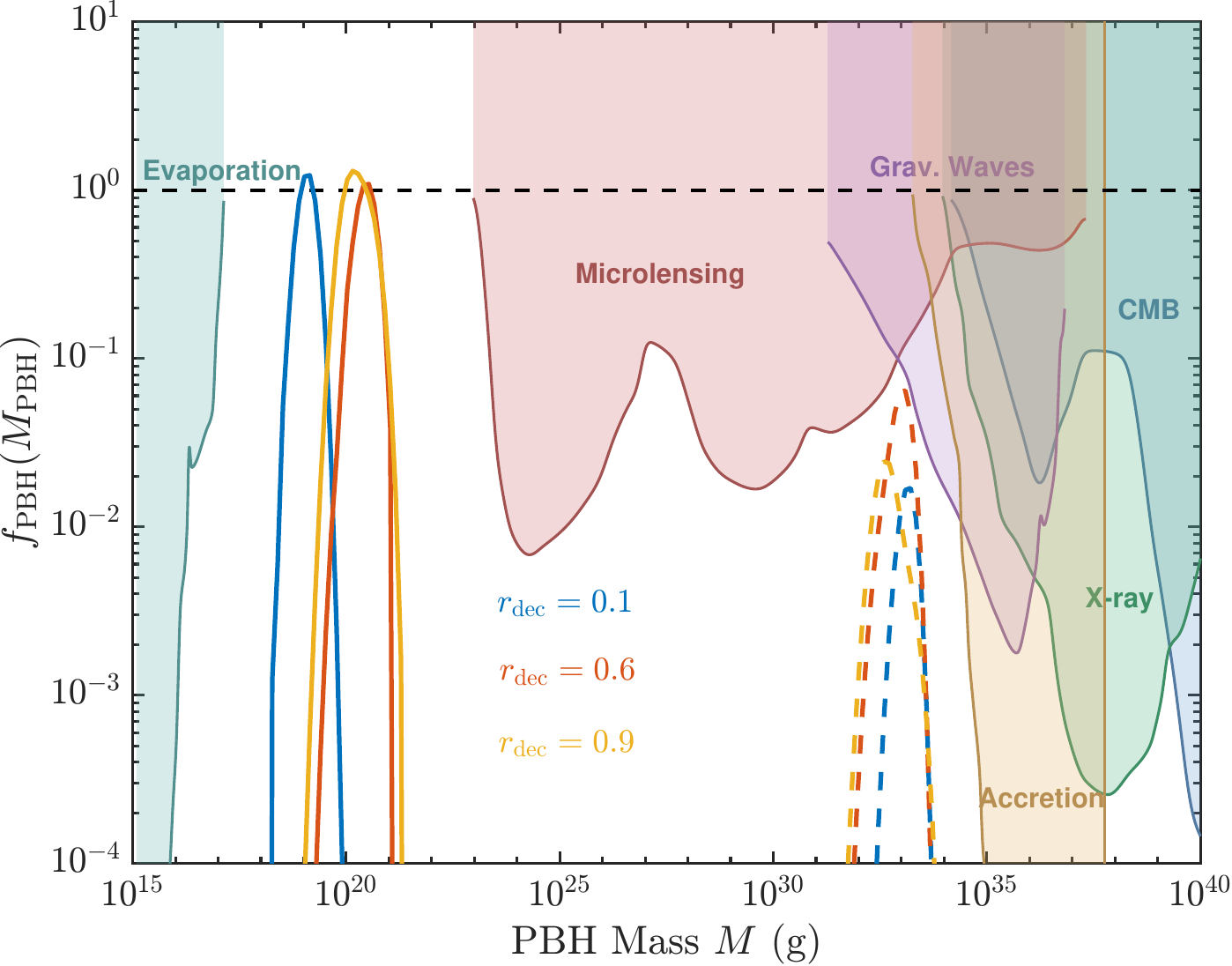}  
        \caption{Peaked spectrum}
    \end{subfigure}
    \hfill
    \begin{subfigure}[b]{0.7\textwidth}
        \centering
        \includegraphics[width=\textwidth]{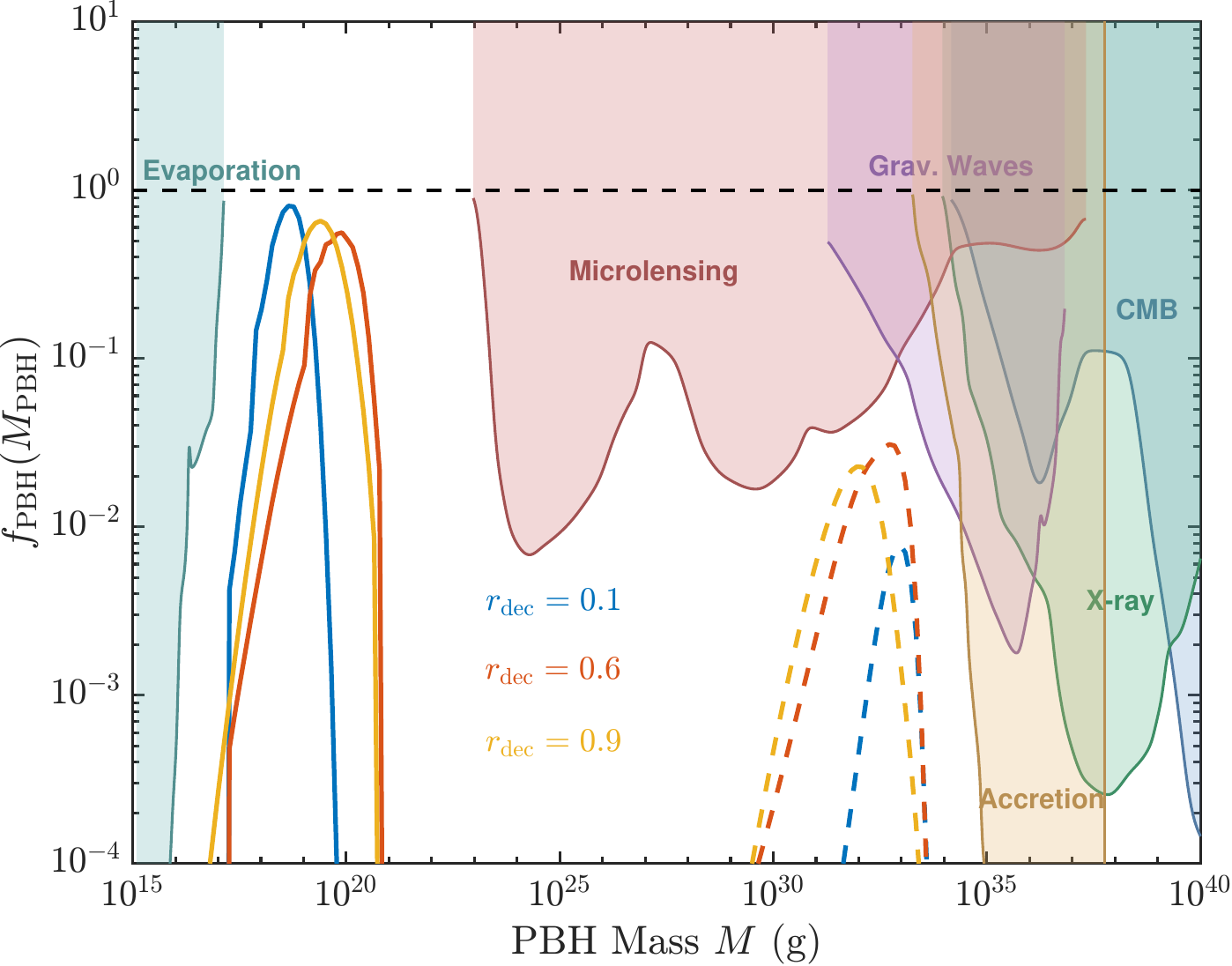} 
        \caption{Broad spectrum}
    \end{subfigure}   
    \caption{The fractional abundance of PBHs for the parameter sets in Table a) \ref{tab:pbh_parameters} and b) \ref{tab:pbh_parametersb}. $f_{\mathrm{PBH}}$ as a function of the PBH mass, computed by incorporating the full non-linear (all-orders) non-Gaussianity. Solid lines (Group 1) produce asteroid-mass PBHs reaching $f_{\mathrm{PBH}} \approx 1$. Dashed lines (Group 2) target solar-mass PBHs ($M \sim M_{\odot}$). The shaded  regions are excluded by current observational constraints, including evaporation\cite{Clark:2016nst,Mittal:2021egv,Laha:2020ivk,Berteaud:2022tws,Boudaud:2018hqb,DeRocco:2019fjq,Carr:2009jm}, microlensing\cite{Niikura:2017zjd,EROS-2:2006ryy,Niikura:2019kqi,Oguri:2017ock}, CMB\cite{Serpico:2020ehh}, gravitational waves from merging events\cite{Green:2020jor}, and X-ray accretion limits\cite{Piga:2022ysp}. See \cite{Carr:2025kdk,Carr:2020gox} for detailed reviews.
} 
    \label{fig:PBHlisa}
\end{figure}

These significantly enhanced curvature perturbations upon horizon reentry lead to the formation of PBHs. Fig.~\ref{fig:PBHlisa} displays the resulting fractional PBH abundance, $f_{\mathrm{PBH}}$, with the upper panel for a peaked spectrum while the lower panel for a broad spectrum. Our calculation incorporates the full non-linear non-Gaussianity of the curvature perturbations, and this non-Gaussian tail inherently changes the PBH production compared to the purely Gaussian regime, allowing Group 1 to successfully account for 100\% of the dark matter. Group 2 targets the solar-mass range, which could lead to observable PBH signals at LIGO/Virgo.

In order for the PBH to be the entirety of dark matter, it is crucial that $c_2 \leq 0$. If $c_2>0$, the radial direction stops rolling at $f$, and the curvature perturbations at high $k$ have a flat spectrum. The PBH spectrum is still peaked, with the high-mass suppression given by the suppressed axion perturbations at low $k$ and the low-mass suppression  given by smaller horizon masses at higher temperature. However, the low-mass slope is $\Omega_{\rm PBH}(M)\propto M^{-1/2}$, and the evaporation bound is violated. In the literature, this constraint in the axion curvaton scenario is avoided by  $T_{\rm dec} \lesssim 10^6$ GeV so that the curvature perturbations at high $k$ is suppressed by the lower energy fraction of the curvaton when those modes enter the horizon. 
%\clearpage

\begin{figure}[tbp]
    \centering
    \includegraphics[width=0.85\linewidth]{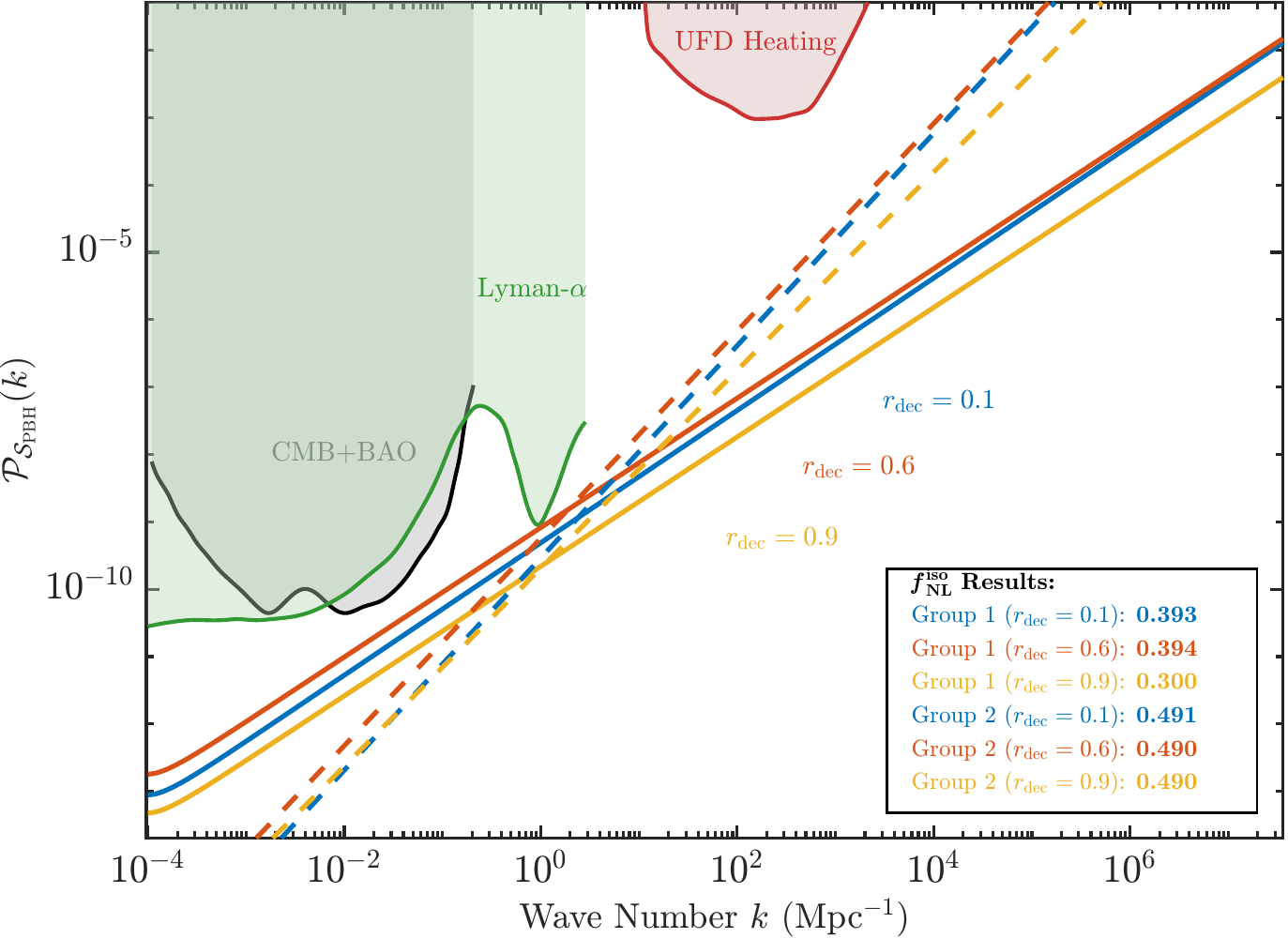}
    \caption{ The primordial black hole dark matter isocurvature power spectrum $\mathcal{P}_{\mathcal{S}_{\mathrm{PBH}}}(k)$ induced by the angular fluctuations of the curvaton field based on Table.~\ref{tab:pbh_parameters}. Solid lines represent Group 1 and dashed lines represent Group 2. The shaded regions show the observational upper limits on isocurvature perturbations, properly accounting for the recent CMB+BAO and Lyman-$\alpha$ constraints derived in the literature\cite{Buckley:2025zgh}. The red shaded region is constrained by ultra-faint dwarf galaxies\cite{Graham:2024hah}. The non-Gaussianity parameters for the isocurvature mode $f_{\mathrm{NL}}^{\mathrm{iso}}$ are presented in the bottom right.}
    \label{fig:PBHisolisa}
\end{figure}

A critical challenge for curvaton-induced PBH models arises from the constraints on isocurvature perturbations. As shown in Fig.~\ref{fig:PBHisolisa}, the dynamical evolution of the radial field in our model naturally induces a blue tilt in the isocurvature power spectrum $\mathcal{P}_{\mathcal{S}_{\mathrm{PBH}}}(k)$. We verify that our chosen parameter sets in Table~\ref{tab:pbh_parameters}
sufficiently suppress the large-scale angular fluctuations to
satisfy the limits set by CMB+BAO measurements as well as the recently updated Lyman-$\alpha$ forest bounds \cite{Buckley:2025zgh}. Similarly, parameter sets in Table~\ref{tab:pbh_parametersb} also satisfy the constraints above. The non-Gaussianity parameter of the isocurvature perturbations is also shown in the box inside the figure, which matches well with the analytical estimation in Eq.~\eqref{eq:fNLDM}.

\begin{figure}[t!]
    \centering
    \begin{subfigure}[b]{0.8\textwidth}
        \centering
        \includegraphics[width=\textwidth]{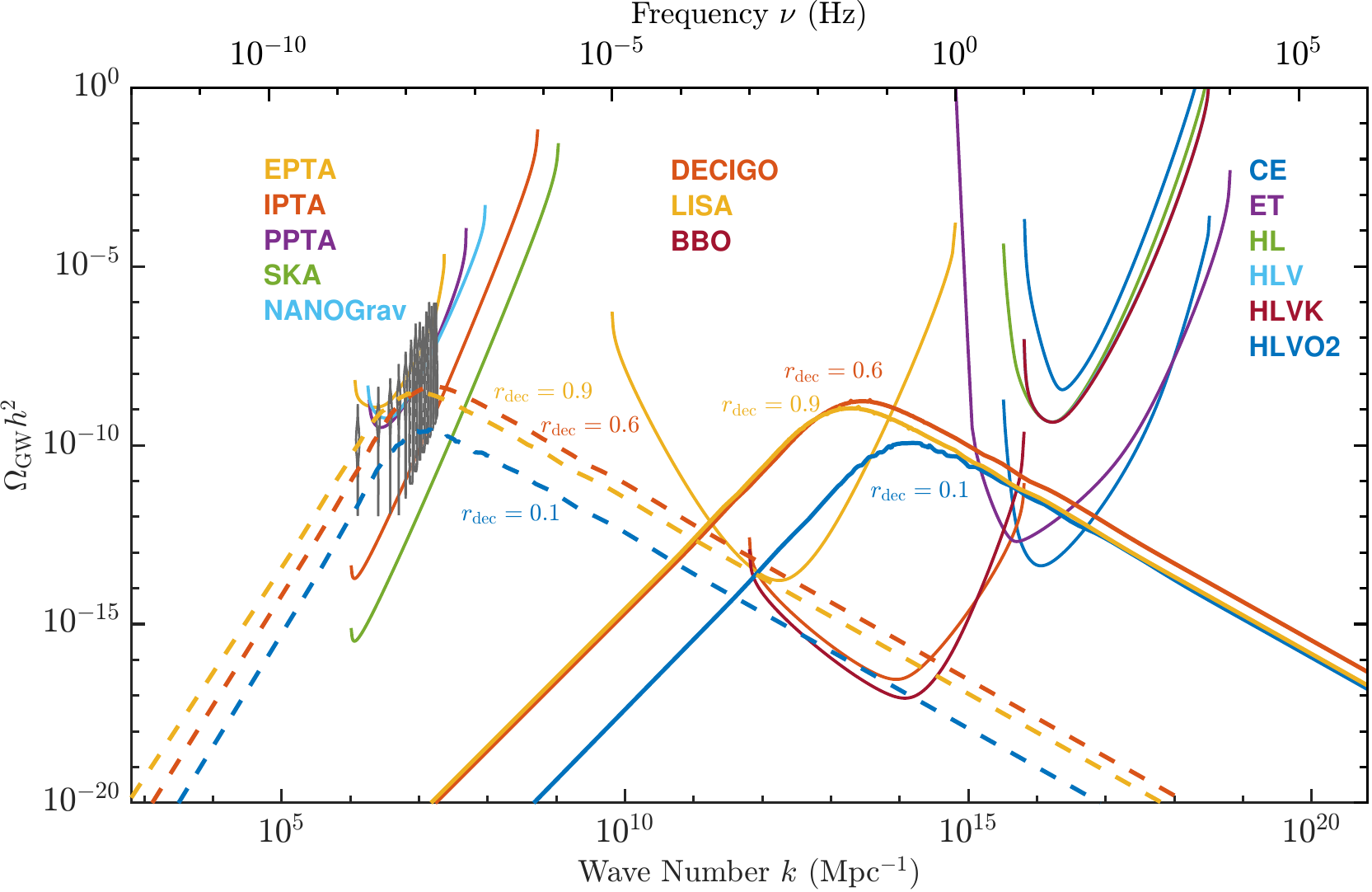}  
        \caption{Peaked power spectrum}
    \end{subfigure}
    \hfill
    \begin{subfigure}[b]{0.8\textwidth}
        \centering
        \includegraphics[width=\textwidth]{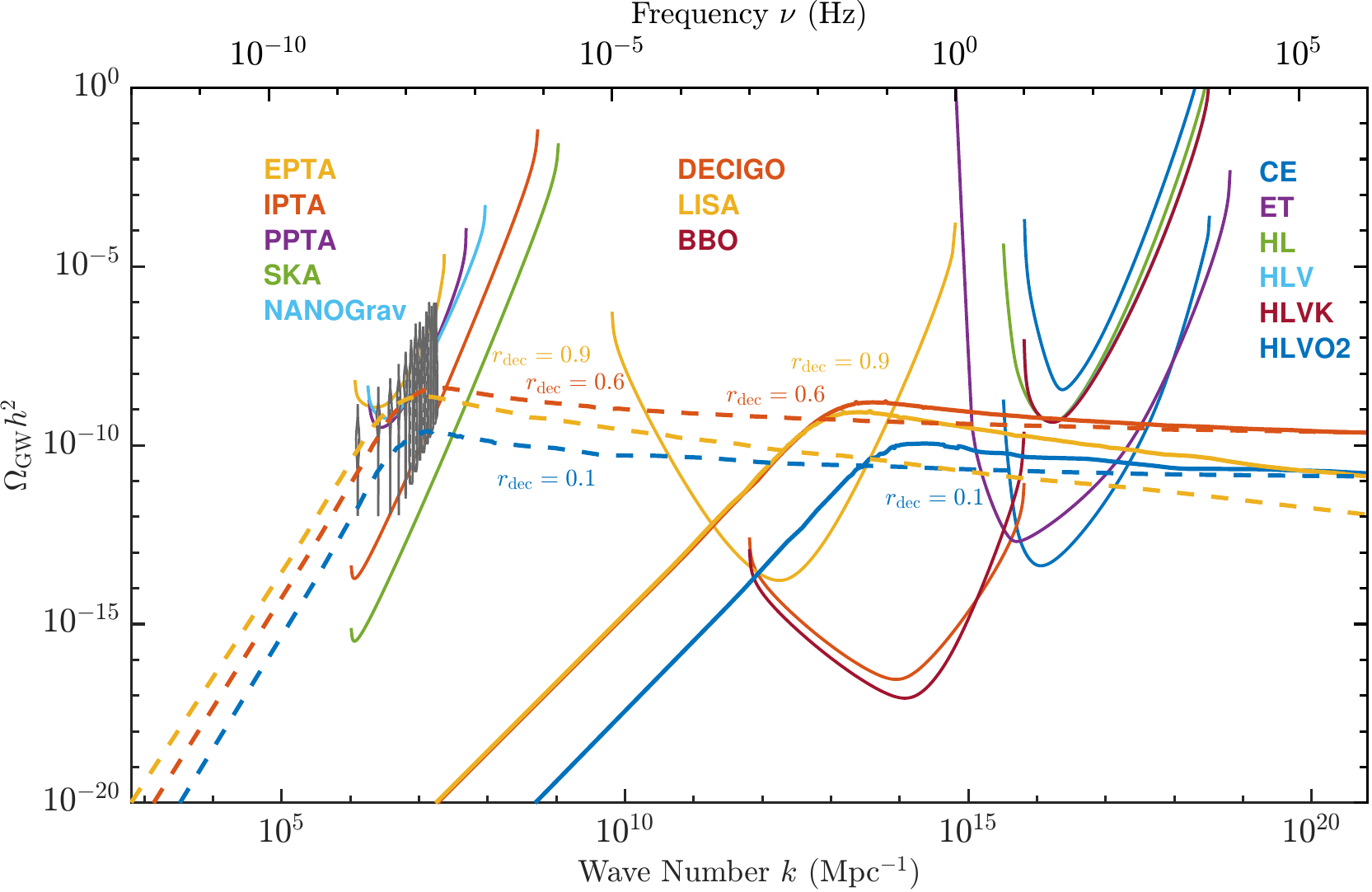} 
        \caption{Broad power spectrum}
    \end{subfigure}   
    \caption{The energy density spectrum of the scalar-induced gravitational wave background, $\Omega_{\mathrm{GW}}h^2$, for the parameter sets in Table a) \ref{tab:pbh_parameters} and b) \ref{tab:pbh_parametersb}, plotted against the comoving wavenumber $k$ and frequency $\nu$. Group 1 signals (solid lines) peak in the frequency range detectable by LISA\cite{LISA:2022kgy}, DECIGO/BBO\cite{Yagi:2011wg}, ET\cite{Punturo:2010zz,Hild:2010id,Sathyaprakash:2012jk,ET:2019dnz}, and CE\cite{LIGOScientific:2016wof,Reitze:2019iox}. Group 2 signals (dashed lines) peak in the nanohertz range, relevant for PTAs (NANOGrav\cite{McLaughlin:2013ira,NANOGRAV:2018hou,Aggarwal:2018mgp,Brazier:2019mmu}, EPTA\cite{Kramer:2013kea,EPTA:2015qep,EPTA:2015gke}, PPTA\cite{Manchester:2012za}), and SKA\cite{Carilli:2004nx,Janssen:2014dka,Weltman:2018zrl}. We also plot aLIGO\cite{Harry:2010zz,LIGOScientific:2014pky}, aVirgo\cite{VIRGO:2014yos}, and the Kamioka Gravitational-Wave Detector (KAGRA)\cite{Somiya:2011np,Aso:2013eba,KAGRA:2018plz,KAGRA:2019htd} for reference. See \cite{Schmitz:2020syl} for a detailed review.}
    \label{fig:GWlisa}
\end{figure}

%\clearpage
Accompanying these scalar perturbations are the SIGWs generated during the radiation-dominated era. In Fig.~\ref{fig:GWlisa}, we present the energy density parameter $\Omega_{\mathrm{GW}}h^2$. The resulting GW spectra inherit the peaked nature of their parent scalar spectra for the upper panel. Group 1 produces a striking signal in the millihertz to hertz range, making it a prime target for future space-based interferometers like LISA, DECIGO, and BBO. Conversely, Group 2 yields a signal firmly in the nanohertz band, aligning seamlessly with the detection ranges of Pulsar Timing Arrays (PTAs) and SKA. Notably, in the lower panel, the broad SIGW signal spanning from nanohertz to millihertz frequencies indicates that this specific scenario could be jointly probed by both PTA collaborations and future multi-band observatories such as LISA, ET, and CE simultaneously.

\begin{figure}[t!]
    \centering
    \begin{subfigure}[b]{0.9\textwidth}
        \centering
        \includegraphics[width=\textwidth]{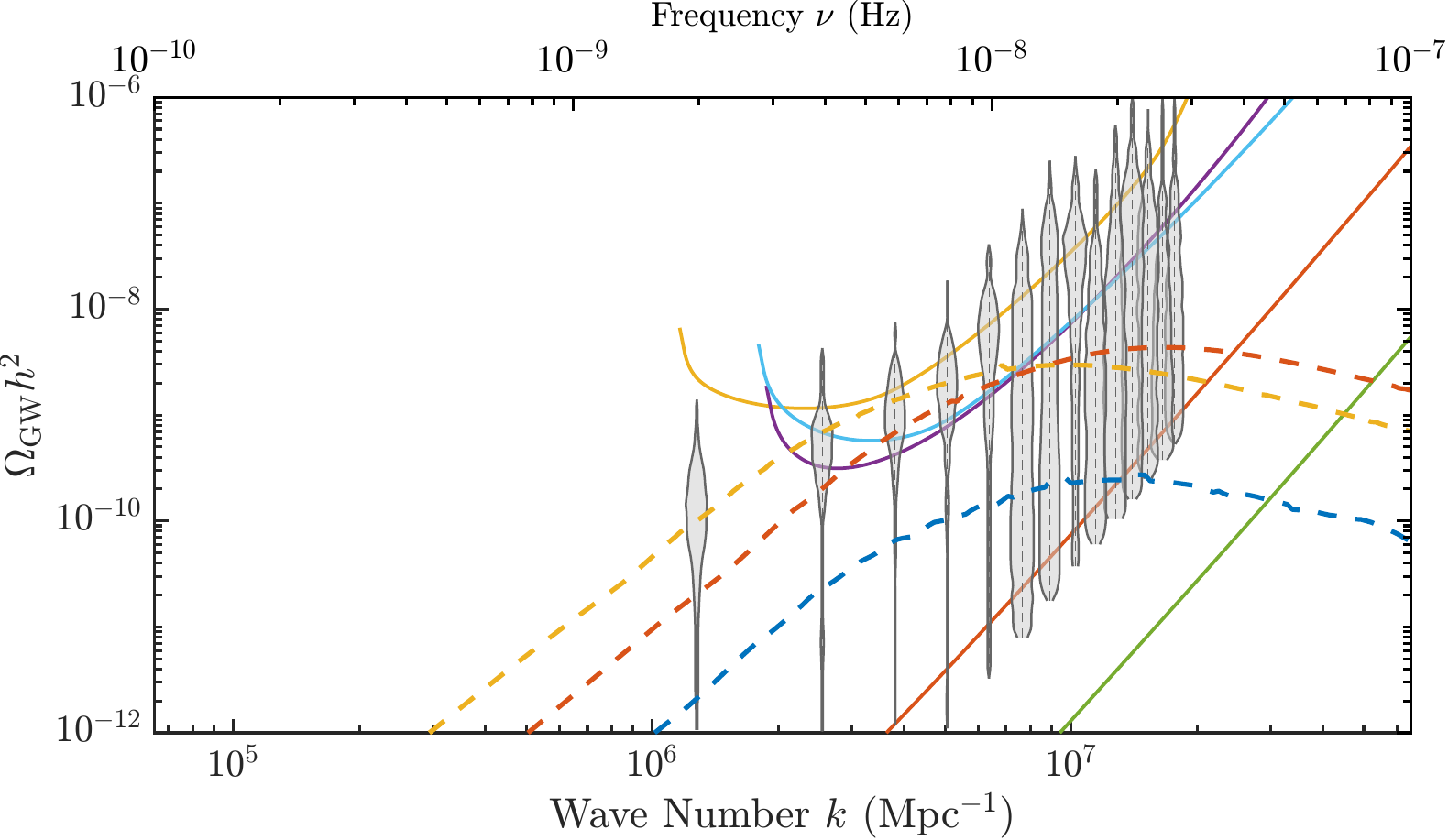}  
        \caption{Peaked power spectrum}
    \end{subfigure}
    \hfill
    \begin{subfigure}[b]{0.9\textwidth}
        \centering
        \includegraphics[width=\textwidth]{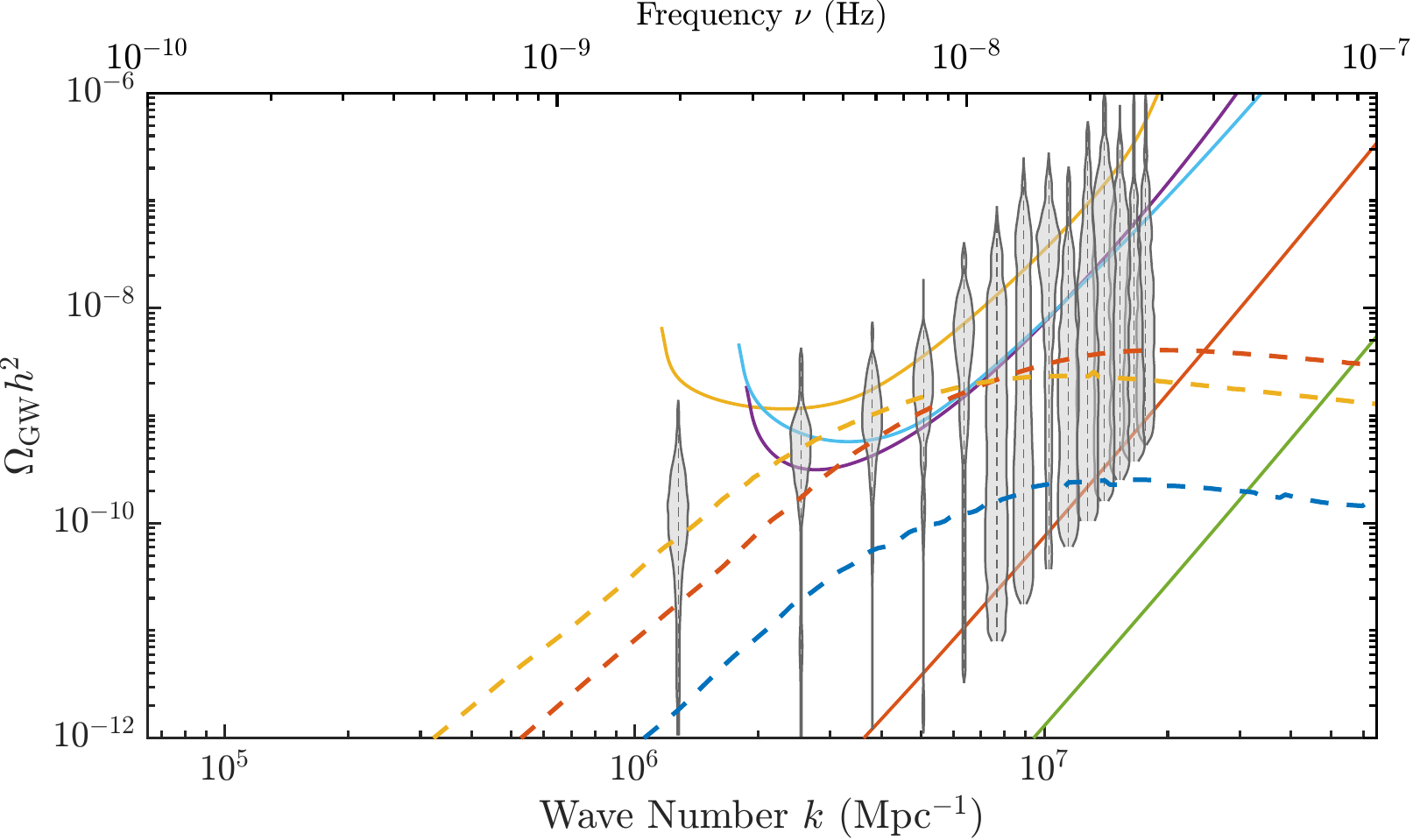} 
        \caption{Broad power spectrum}
    \end{subfigure}   
  \caption{
        A zoomed-in view of the nanohertz frequency band from Fig.~\ref{fig:GWlisa}. 
        The dashed curves correspond to the energy density spectra of the scalar-induced gravitational waves predicted by our Group 2 parameter sets. 
        The violin plots display the posterior probability distributions (free spectrum) of the stochastic gravitational wave background amplitude extracted from the NANOGrav dataset. 
        This detailed view highlights the capability of the axion curvaton model to accommodate the recent pulsar timing array observations.
    }
    \label{fig:Zoomin}
\end{figure}

To explicitly demonstrate the alignment, Fig.~\ref{fig:Zoomin} provides a zoomed-in view of the nanohertz frequency band, comparing our theoretical predictions with recent PTA observations. The violin plots represent the posterior probability distributions of the free spectrum for the stochastic GW background (SGWB), as extracted from the NANOGrav 15-year data set\cite{NANOGrav:2023hvm}. Overlaid on these data points are the dashed curves representing the SIGW spectra predicted by our Group 2 benchmark scenarios. As clearly shown, the theoretical spectra safely traverse the high-probability regions of the NANOGrav measurements. This compelling agreement indicates that the exact same cosmological dynamics responsible for producing solar-mass PBHs as a viable dark matter candidate can simultaneously offer a natural, primordial origin for the SGWB signals recently reported by PTA collaborations.

\clearpage

\subsection{Correlation between GW signal and DM isocurvature}

One of the most intriguing phenomenological features of the axion curvaton model is the intrinsic correlation between the small-scale GW
signals and the large-scale dark matter isocurvature perturbations. The IR tail of the curvature power spectrum is well approximated by a power-law $\mathcal{P}_{\zeta}(k) \propto k^{2\lambda_1}$ for $k \ll k_{\mathrm{peak}}$, which consequently dictates the IR slope of the induced stochastic GW background, $n_{\mathrm{GW,IR}} = 4\lambda_1$.%
\footnote{The mapping between the scalar tilt $2\lambda_1$ and $n_{\mathrm{GW,IR}}$ is dictated by the dominant integration regime. In the local-mode dominated regime ($2\lambda_1 \le 1.5$), the scalar IR tail is relatively flat, and IR GWs are sourced directly by the nonlinear conversion of large-scale scalar perturbations upon horizon reentry. The local integral cutoff dictates the overall behavior, yielding a purely quadratic mapping $n_{\mathrm{GW,IR}} = 4\lambda_1$. In the peak-mode dominated regime ($2\lambda_1 > 1.5$), the scalar IR tail is extremely steep. Large-scale scalar fluctuations are negligible, and deep IR GWs are generated almost entirely by intense short-wavelength fluctuations at $k_{\mathrm{peak}}$ leaking into the IR via the momentum difference $\vec{p} - \vec{q} = \vec{k}$. The integral loses dependence on the local scalar tilt and collapses to the theoretical upper bound allowed by causality, locking the IR GW slope at $n_{\mathrm{GW,IR}} = 3$ (up to a long-wavelength renormalization logarithmic correction $k^3 \ln^2 k$).}
Meanwhile, this identical long tail extends to CMB and intermediate scales, sourcing the PBH dark matter isocurvature perturbations.

\begin{figure}[t]
    \centering
    \includegraphics[width=1\linewidth]{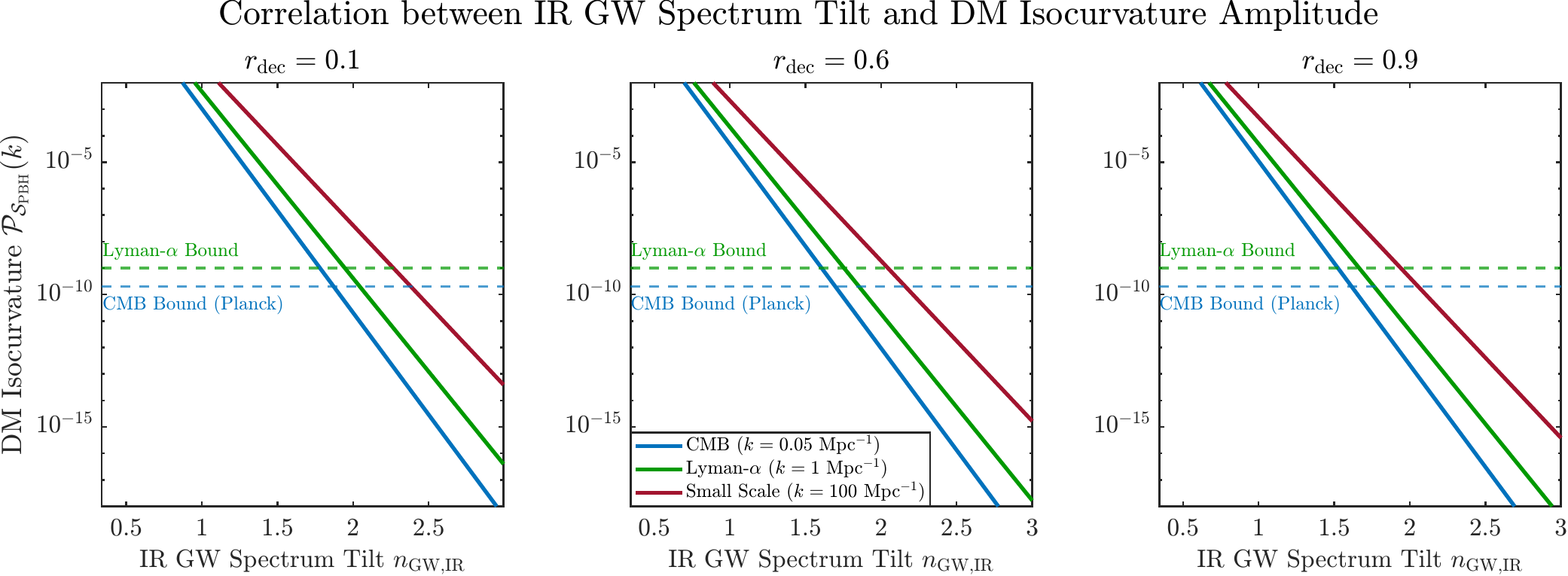}
    \caption{Correlation between the infrared (IR) tilt of the GW power spectrum, $n_{\mathrm{GW,IR}}$, and the amplitude of dark matter isocurvature perturbations, $\mathcal{P}_{\mathcal{S}_{\mathrm{PBH}}}(k)$. Because the IR slope of the GW spectrum is directly determined by the scalar tilt via $n_{\mathrm{GW,IR}} = 4\lambda_1$ (for $k \ll k_{\mathrm{peak}}$, $n_{\mathrm{GW,IR}}<3$), this plot explicitly bridges the GW observables with CMB-scale constraints.
        We focus on a benchmark scenario where PBHs constitute the entirety of dark matter ($f_{\mathrm{PBH}} \approx 1$) by anchoring the peak of the curvature power spectrum at $k_{\mathrm{peak}} \sim 10^{13} \text{ Mpc}^{-1}$ with an amplitude $\mathcal{P}_{\zeta}(k_{\mathrm{peak}}) \sim 0.01$. 
        The three panels represent different curvaton decay fractions: $r_{\mathrm{dec}} = 0.1, 0.6$, and $0.9$. In each panel, the colored solid curves denote the isocurvature amplitude evaluated at three representative scales: the CMB scale ($k=0.05 \text{ Mpc}^{-1}$, blue), the Lyman-$\alpha$ scale ($k=1 \text{ Mpc}^{-1}$, green), and a smaller scale ($k=100 \text{ Mpc}^{-1}$, red) targeted by near-future observations. The horizontal blue and green dashed lines indicate the upper bound on isocurvature perturbations from Planck CMB measurements and Lyman-$\alpha$ at each scale; see Fig~\ref{fig:PBHisolisa}. Here the scales of the CMB and Lyman-$\alpha$ are chosen so that the constraint is the strongest. 
        %Both Lyman-$\alpha$ and CMB give comparable results around $k\approx 10^{-4}\text{ Mpc}^{-1}\sim10^{-2}\text{ Mpc}^{-1}$. Lyman-$\alpha$ additionally has a bound of $10^{-9}$ around $k\approx 1\text{ Mpc}^{-1}$ which is also relevant.
    }
    \label{fig:correlation_tilt_iso}
\end{figure}

To explicitly demonstrate this correlation, we present Fig.~\ref{fig:correlation_tilt_iso}. In this analysis, we impose the condition that PBHs constitute all of dark matter ($f_{\mathrm{PBH}} \approx 1$) with a specific target mass. This effectively anchors both the position ($k_{\mathrm{peak}} \sim 10^{13} \text{ Mpc}^{-1}$) and the amplitude ($\mathcal{P}_{\zeta} \sim 0.01$) of the spectrum peak. We then continuously vary the Hubble-induced mass parameter $c_1$ to scan the IR tilt $2\lambda_1$. To compensate for the variation in the rolling velocity and preserve the anchored peak, one could simultaneously adjust the initial field value $\varphi_p$ (in the sub-Planckian regime for slower rolling) or proportionally rescale the scale $f$ and $H$ (for faster rolling to avert quantum diffusion). 

The results in Fig.~\ref{fig:correlation_tilt_iso} reveal a stringent anti-correlation:~a shallower IR tilt (smaller $2\lambda_1$) implies that the power spectrum decays less towards larger scales, thereby resulting in  enhanced isocurvature perturbations at the CMB ($k=0.05 \text{ Mpc}^{-1}$) and Lyman-$\alpha$ ($k=1 \text{ Mpc}^{-1}$) scales. Current constraints from Planck and Lyman-$\alpha$ already place a lower bound on the IR tilt, ruling out scenarios with excessively flat IR tails.

More importantly, this correlation opens a unique multi-messenger pathway to test the model. If a stochastic GW background is detected by forthcoming interferometers (e.g., LISA or DECIGO), we may measure its IR spectral tilt $n_{\mathrm{GW,IR}}$.
%will break the parameter degeneracies in the inflationary potential.
Once the tilt is pinned down, our model predicts the amplitude of dark matter isocurvature perturbations. As illustrated by the red curves in Fig.~\ref{fig:correlation_tilt_iso}, intermediate scales such as $k = 100 \, \mathrm{Mpc}^{-1}$, which are prime targets for near-future 21-cm observations~\cite{deKruijf:2024voc}, possess the highest sensitivity and are poised to probe the parameter space with steeper slopes that currently evade the Planck limits. (CMB spectral distortions may provide integral constraints, but their sensitivity to this specific isocurvature mode is limited; see, e.g.,~\cite{Lee:2021bmn}.)

%\clearpage

\section{Conclusions}
\label{sec:summary}
In this paper, we have shown that an axion field can develop field fluctuations peaked at a characteristic wavenumber. This arises through the time evolution  of the radial direction of a symmetry-breaking field during inflation, driven by Hubble-induced masses. For a particular combination of the signs of the Hubble-induced masses, the axion decay constant starts from a large value, decreases, and then increases during inflation. The peak of the axion fluctuation spectrum is set by the moment when the decay constant reaches its minimum, while the IR and UV slopes are controlled by the magnitudes of the Hubble-induced masses.

These axion field perturbations can be converted into curvature perturbations via the curvaton mechanism. We computed the resulting GW and PBH abundances. The GW spectrum has a triangular shape, with IR and UV slopes determined by the Hubble-induced masses.
The NANOGrav signal can be accommodated by our mechanism without overproducing PBHs, once the non-Gaussianity of the curvature perturbations is properly accounted for. If the GW spectrum is measured with greater precision in the future, the model parameters will be further constrained and GW signals at higher frequencies will be predicted.

The PBHs produced in our setup can account for all of dark matter. Regardless of whether they constitute all of dark matter, PBHs in our model generically carry blue-tilted isocurvature perturbations on large scales. CMB constraints require the axion field perturbations to be sufficiently suppressed in the IR, which in turn requires the IR slope of the GW spectrum to be sufficiently steep.

It is worth comparing our mechanism with existing axion curvaton models~\cite{Kawasaki:2012wr,Ando:2017veq,Inomata:2020xad,Ferrante:2023bgz}. While these models share the phenomenological success of producing PBHs that can explain dark matter or LIGO events, together with broad and detectable stochastic GW backgrounds, their underlying inflationary dynamics and resulting power spectra differ significantly from ours. Previous studies typically consider scenarios in which the radial field has only a positive Hubble-induced mass and stops evolving once it reaches its minimum. Depending on whether perturbations re-enter the horizon before or after curvaton decay, these models predict fixed UV slopes for the curvature power spectrum (e.g., $\mathcal{P}_{\zeta} \propto k^{-2}$), often accompanied by a broad flat plateau. In contrast, our model introduces a negative Hubble-induced mass ($c_2 < 0$) that drives the radial field to roll continuously. As a result, a flat plateau is generically absent, and both IR and UV slopes of the triangular spectrum are not fixed but are dynamically determined by the Hubble-induced masses ($c_1$ and $c_2$), making the spectral shape considerably more flexible.

Our setup opens a new avenue for studying PBH formation in highly non-Gaussian regimes. While our full-order $\delta N$ mapping successfully evaluates curvature perturbations within the valid quadratic potential regime ($\theta \lesssim 0.3$), extending this treatment to the full cosine potential remains an important task. In the non-perturbative regime of large initial misalignment angle ($\theta \sim 1$), the full curvature perturbation $\zeta$ deviates significantly from its Gaussian component $\zeta_{\rm G}$ due to the periodicity of the potential. Since standard methods for computing PBH abundance rely on collapse thresholds derived under nearly Gaussian assumptions, applying them in this strongly non-linear regime can introduce substantial systematic errors. A robust computation of the collapse threshold using the full curvature perturbation rather than only its Gaussian part will therefore be essential for accurate treatments of non-polynomial curvaton potentials.

Finally, while we focused on adiabatic perturbations, the peaked axion field fluctuations can also source isocurvature perturbations. In particular, when axion dark matter is produced by the (kinetic) misalignment mechanism, or when baryon asymmetry is generated by axiogenesis, the axion field perturbations source dark matter or baryon isocurvature perturbations that inherit the peaked spectrum. Such peaked isocurvature perturbations could serve as a distinctive target for future galaxy surveys.

\section*{Acknowledgment}
We thank Yunjia Bao, Lorenzo Calibbi, Dhong Yeon Cheong, Gabriele Franciolini,  Masahiro Kawasaki, Hanwen Tai, and Isaac R.~Wang for useful discussions.
This work was supported by the Department of Energy grant DE-SC0009924 [KH and LTW] and the World Premier International Research Center Initiative (WPI), MEXT, Japan (Kavli IPMU) [KH]. JY acknowledges the School of Physics at Nankai University for financial support. 

\appendix

\section{PBH full order computation}
\label{app:PBH}
 On super-horizon scales, using the gradient expansion expression in Eq.~\eqref{eq:deltanl} and assuming spherical symmetry, the compaction function becomes time-independent and can be written entirely in terms of the curvature perturbation:
\begin{equation}
\mathcal{C}(r) = -2\Xi \, r \zeta'(r)
\Bigl[ 1 + \frac{r}{2} \zeta'(r) \Bigr]
= \mathcal{C}_1(r) - \frac{1}{4\Xi} \mathcal{C}_1(r)^2,
\qquad
\mathcal{C}_1(r) \equiv -2\Xi \, r \zeta'(r).
\label{eq:Cfromzeta}
\end{equation}
Here $\mathcal{C}_1$ denotes the linear piece of the compaction. The scale $r_m$ is defined as the location where $\mathcal{C}(r)$ is maximised, satisfying $\mathcal{C}'(r_m)=0$.

The scale $r_m$ is related to the comoving wavenumber $k$ of the perturbation via a dimensionless factor $\kappa$ that depends on the shape of the power spectrum:
\begin{equation}
r_m k = \kappa .
\label{eq:rmk}
\end{equation}
The dimensionless factor $\kappa$ depends functionally on the shape of the power spectrum; its value typically increases as the spectrum becomes broader. For instance, in the case of a step-function (broad) power spectrum, $\kappa$ approaches a constant value (e.g., $\kappa \simeq 4.49$ \cite{Musco:2020jjb}). Similarly, the threshold for collapse $\mathcal{C}_{\rm th}$ (or, equivalently, the density contrast threshold $\delta_c$) is also sensitive to the power spectrum shape, for broad spectrum $\mathcal{C}_{\rm th}=\delta_c=0.56$ \cite{Musco:2020jjb}. In this work, we adopt the threshold and $\kappa$ values obtained for a broad power spectrum. This choice is motivated by two observations. First, we have numerically verified that the dependence of the final PBH abundance on the precise value of $\kappa$ is mild: varying $\kappa$ over the range expected for different spectral shapes results only in an $\mathcal{O}(1)$ change in the computed mass fraction. Second, and more importantly, the threshold $\delta_c$ has a more significant impact on the abundance; our own power spectrum, when compared with both log-normal spectra of large width and the step-function broad spectrum, exhibits a similar two-dimensional integration contour in the $(\mathcal{C}_{\rm G},\zeta_{\rm G})$ plane(see eg.\cite{Ferrante:2022mui}), justifying our use of the broad-spectrum benchmark. It is important to emphasize that both $\kappa$ and $\delta_c$ are conventionally extracted from numerical simulations that assume purely Gaussian curvature perturbations; consequently, they serve as reference values only in the absence of primordial non-Gaussianity (NG). In the presence of significant primordial NG, these thresholds may receive corrections, but a consistent inclusion of such effects is beyond the scope of the present work and is left for future study.

When primordial NG is present, these thresholds can in principle be modified. Recent works have begun to explore such corrections within the framework of quadratic or cubic approximations \cite{Escriva:2022pnz,Kehagias:2019eil}. However, a full all-orders computation—particularly for models with non-polynomial potentials such as the cosine potential of an axion-like curvaton—remains an open and necessary direction for future study. In this work, we focus on the parameter regime ($\theta \lesssim 0.3$) where the curvaton field exhibits a quadratic potential, and we therefore adopt the values of $\kappa$ and $\delta_c$ obtained under the Gaussian assumption \cite{Musco:2020jjb,Germani:2018jgr}. We acknowledge that this choice may introduce an $\mathcal{O}(1)$ systematic effect in the predicted abundance when primordial NG is large, but a consistent inclusion of NG in the threshold calculation is beyond the scope of the present analysis.

When primordial NGs are present, $\zeta = \mathcal{F}(\zeta_{\rm G})$, the linear compaction becomes
\begin{equation}
\mathcal{C}_1(r) = \mathcal{C}_{\rm G}(r) \,
\frac{d\mathcal{F}}{d\zeta_{\rm G}}, \qquad
\mathcal{C}_{\rm G}(r) \equiv -2\Xi \, r \zeta_{\rm G}'(r),
\label{eq:C1NG}
\end{equation}
and the full compaction reads
\begin{equation}
\mathcal{C}(r) = \mathcal{C}_{\rm G}(r) \,
\frac{d\mathcal{F}}{d\zeta_{\rm G}}
- \frac{1}{4\Xi} \mathcal{C}_{\rm G}(r)^2
\left(\frac{d\mathcal{F}}{d\zeta_{\rm G}}\right)^2.
\label{eq:CfullNG}
\end{equation}
The Gaussian variables $\zeta_{\rm G}$ and $\mathcal{C}_{\rm G}$ are correlated; their joint probability distribution function (PDF) is given by
\begin{equation}
{\rm P}_{\rm G}(\mathcal{C}_{\rm G},\zeta_{\rm G}) =
\frac{1}{2\pi\sigma_c\sigma_r\sqrt{1-\gamma_{cr}^2}}
\exp\!\left[-\frac{\zeta_{\rm G}^2}{2\sigma_r^2}\right]
\exp\!\left[-\frac{1}{2(1-\gamma_{cr}^2)}
\left(\frac{\mathcal{C}_{\rm G}}{\sigma_c}
- \frac{\gamma_{cr}\zeta_{\rm G}}{\sigma_r}\right)^2\right],
\label{eq:jointPDF}
\end{equation}
where the variances and correlation coefficient are
\begin{align}
\sigma_c^2 &\equiv \langle\mathcal{C}_{\rm G}\mathcal{C}_{\rm G}\rangle
= \frac{4\Xi^2}{9} \int_0^\infty \frac{dk}{k} \,
(k r_m)^4 W^2(k,r_m) T^2(k,r_m) \mathcal{P}_{\zeta}(k), \\
\sigma_{cr}^2 &\equiv \langle\mathcal{C}_{\rm G}\zeta_{\rm G}\rangle
= \frac{2\Xi}{3} \int_0^\infty \frac{dk}{k} \,
(k r_m)^2 W(k,r_m) W_s(k,r_m) T^2(k,r_m) \mathcal{P}_{\zeta}(k), \\
\sigma_r^2 &\equiv \langle\zeta_{\rm G}\zeta_{\rm G}\rangle
= \int_0^\infty \frac{dk}{k} \,
W_s^2(k,r_m) T^2(k,r_m) \mathcal{P}_{\zeta}(k), \\
\gamma_{cr} &\equiv \frac{\sigma_{cr}^2}{\sigma_c \sigma_r}.
\end{align}
Here $\mathcal{P}_{\zeta}(k)$ is the power spectrum of the Gaussian field $\zeta_{\rm G}$, $W(k,R)$ is a top-hat window function smoothing over a scale $R=r_m$, $W_s(k,r)=\sin(kr)/(kr)$, and $T(k,\tau)$ is the linear transfer function that accounts for sub-horizon damping\cite{Musco:2020jjb}.

\bibliographystyle{JHEP}
\bibliography{PACM}

\end{document}